# Cadmium Resonance Parameters from Neutron Capture and Transmission Measurements at the RPI LINAC

G. Leinweber[1], D. P. Barry[1], R. C. Block[1], J. A. Burke[1], K. E. Remley[1], M. J. Rapp[1], and Y. Danon[2]

[1] Bechtel Marine Propulsion Corporation, P.O. Box 1072, Schenectady, NY 12301-1072
[2] Rensselaer Polytechnic Institute, 110 8th St. Troy, NY 12180

Corresponding Author: Dr. Gregory Leinweber (518) 276-4006  email: leinwg@rpi.edu

Cadmium has been used historically as an important component of integral experiments because of its high thermal neutron absorption cross section. Correct interpretation of such experiments depends on accurate differential neutron cross section measurements. The 60 MeV electron accelerator at the Gaerttner LINAC Center was used to generate neutrons for neutron capture and total cross section measurements of natural Cd. Measurements were performed in the thermal and epithermal resonance range with sample thicknesses ranging from $1 \times 10^{-4}$ to $4 \times 10^{-2}$ atoms per barn. A full resonance region analysis was performed in order to determine the thermal cross sections and resonance integrals of the cadmium isotopes. The Bayesian R-Matrix code SAMMY 8.0 was used to shape fit the data and extract the resonance parameters. Resonance parameter and cross section uncertainties were determined from their primary components: transmission background, capture normalization, experimental resolution function, burst width, sample thickness, and counting statistics. The experiments were analyzed for consistency within the measured capture and transmission using multiple sample thicknesses. Results are compared to previously published measurements and evaluated nuclear libraries. No major changes to the thermal cross section or the first resonance in Cd113 were identified from the consensus achieved from measurements and evaluations over the past decade.

## I. Introduction

Cadmium is a strong absorber of thermal energy neutrons; the peak cross section of Cd113, 12.2% abundant in natural Cd, is ~20,000 barns. As such it has been used to cover activation foils to quantify the epi-cadmium flux in nuclear criticality benchmarking experiments. It is a material used in neutron activation experiments and wherever the removal of virtually all thermal neutrons is desired. In addition the shape of the cross section on the high energy side of the first resonance is very important to accurate calculation of the Cd cut. Therefore, precise measurements of its thermal cross section shape, and thus the underlying resonance parameters have been undertaken.

The initial suite of natural cadmium transmission and capture measurements were performed in 2002 and 2003. The results were treated as a reference for verifying updated data acquisition equipment in 2007. These equipment-verification data have been included in the current analysis. The consistency of the results have thus been verified not only with samples of multiple thicknesses in independent but overlapping energy regions (thermal and epithermal), but also over a span of years using different time-of-flight (TOF) clocks, analysis codes, and analysts.

A historical review of the results of decades of examination of the prominent 0.178 eV resonance in Cd113 which gives rise to the large thermal cross section is given in Table 1 and Table 2. Table 1 lists the pertinent measurements, and Table 2 gives the most recent evaluations. At the time the current measurements commenced there was some uncertainty in the resonance parameters of the prominent resonance at 0.178 eV in Cd113. However, as a result of the current measurements and those of Kopecky[1], a consensus has emerged.

The experimental methods, data processing methods, and analytical techniques are similar to those used for other measurements from the RPI LINAC for Eu[2], Dy[3], and Re[4]. An abridged description of this measurement was published in Reference 5.

## II. Experimental Conditions

### a. Experimental details

The details of the experiments are given in Table 3 including neutron targets, overlap filters, LINAC pulse repetition rates, and flight path lengths. The nominal resolution, pulse width divided by flight path length was ~2 ns/m for the epithermal experiments. Thermal measurements did not require fine energy resolution. A crucial parameter in each data analysis is the resonance energy at which the transmission background or the capture flux is normalized. This resonance for each measurement is given in Table 3.

The electron linear accelerator at Rensselaer Polytechnic Institute's (RPI) Gaerttner LINAC Center was used to bombard Ta plates in a neutron-producing target. The neutron-producing targets were optimized for each energy range.[6,7,8] Sample placement was automated. All samples were cycled into the neutron beam approximately every hour. In addition to the main data-taking detector, a separate set of neutron detectors operated on an adjacent beamline to measure fluctuations in the neutron beam intensity from sample-to-sample. These data are referred to as beam monitor data and are used to make a small adjustment to the data from the main data-taking detectors.

The zero time for the measured time-of-flight spectrum was determined from a separate measurement of the location of the count rate peak produced by the flash of gamma rays which accompanies each pulse of electrons from the accelerator. This 'gamma flash' coincides with each burst of neutrons from the target.

### b. Sample information

Each experiment subjected a series of sample materials to a neutron flux for the purpose of determining the interaction rate in the sample. Details of the Cd samples and the experiments in which they were used are given in Table 4. Sample thicknesses ranged from approximately $1.1 \times 10^{-4}$ to $4.4 \times 10^{-2}$ atoms per barn. Thick samples were needed to see a significant number of reactions in weak resonances. Thin samples were needed to get optimum transmission for strong resonances. The samples were 50.8 mm in diameter. The uncertainties in sample thicknesses were determined from mass and multiple diameter measurements. The diameter measurements were the primary source of uncertainty in sample thickness. The samples had a nominal purity of 99.99%. A spectrographic analysis of the samples yielded the impurities above 1 ppm as 4 ppm Pb and 3 ppm Au.[9]

c. Capture detector

Capture measurements were made with the same detector for thermal and epithermal energy ranges using the time-of-flight (TOF) method at an approximately 25.5 m flight path. The capture detector is a gamma-ray detector consisting of 20 liters of NaI(Tl) divided into two annular right circular cylinders whose axes were parallel with the neutron beam. Each detector half was divided into 8 pie-shaped, optically-isolated segments.[10] The inside of the detector is lined with a 1.0-cm thick annulus of 99.5 w/o B10 carbide ceramic to reduce the number of scattered neutrons reaching the gamma detector. A total energy deposition of 1 MeV for the epithermal measurement and 2 MeV for the thermal measurement was required to register a capture event. Therefore, the system discriminates against the 478 keV gamma rays from B10(n;α,γ) absorptions. The dead time of the system was set to 1.125 μs. The maximum dead time correction was 8.5%.

d. Transmission detectors

Transmission measurements were made using two separate Li6 glass detectors. Thermal measurements were made at ~15 m using a 7.62-cm (3-in) diameter, 0.3-cm-thick NE 905 Li6 glass scintillation detector (6.6% lithium, enriched to 95% in Li6) coupled to a single photomultiplier. Data from the thermal experiment was analyzed from a few meV to ~10 eV. Cd samples (see Table 4) and empty aluminum sample holders were cycled in and out of the neutron beam.

Epithermal measurements were made at ~25.6 m with a 12.70-cm (5-in.) diameter, 1.27-cm-thick Li6 glass scintillator (with similar Li content and enrichment as the thermal detector) housed in a light-tight aluminum box.[11] Two photomultiplier tubes were positioned out of the neutron beam to collect the scintillation light. Data from the epithermal experiment was analyzed from 15-1000 eV.

The transmission can be expressed as the ratio of the count rate with a sample in the beam to the count rate with samples removed once all count rates have been corrected for background. The transmission can vary strongly with incident neutron energy and is directly linked to the total cross section of the sample material being measured. The maximum dead time correction was 3%.

**III. Data Reduction**
a. Capture Data

Raw capture data were corrected for dead time, normalized to the beam monitors, summed over all cycles, background subtracted, and processed into yields. Yield is defined as the number of neutron captures per neutron incident on the sample. The capture yield, $Y_i$, in time-of-flight channel i, was calculated from Eq. (1).

$$Y_i = \frac{C_i - B_i}{K\left(\phi_i - B_{\phi_i}\right)}, \quad (1)$$

where $C_i$ is the dead-time-corrected and monitor-normalized counting rate of the sample measurement,
$B_i$ is the dead-time-corrected and monitor-normalized background counting rate,
$K$ is the product of the flux normalization factor and efficiency, and
$\phi_i$ is the measured neutron flux shape.
$B_{\phi_i}$ is the dead-time-corrected and monitor-normalized background counting rate in the measured flux shape.

The energy dependent neutron flux shape was measured separately using a 2.54-mm thick, 98.4% enriched $^{10}B_4C$ sample. The total energy deposition discriminator setting was set to a window from 360-600 keV in order to record the 478 keV gamma ray from neutron absorption in B10. To avoid dead-time concerns only two of the 16 NaI segments were active. For the epithermal capture measurement the flux shape measured in this way was corrected for the transmission through the boron carbide sample. The measured flux shape was normalized to a 'black' (saturated) resonance in Cd. The resonance used for this normalization was located at an energy of 0.178 eV for thermal capture or 27 eV for epithermal capture, as shown in the rightmost column of Table 3.

The detector efficiency for a typical gamma cascade was approximately 90%.[3] No provision was made for the detector efficiency in the analysis since the flux was normalized to a black resonance in Cd113. Most of the fitted resonances were from the even-odd nuclei Cd111 and Cd113 with similar binding energies whose efficiencies are expected to be similar. Also, there were no efficiency variations due to resonance energy or spin state in the same isotope because all multiplicities were summed and the total gamma cascade energy did not vary with the resonance energy or spin within the same isotope.

The background in the capture measurements was determined from in-beam measurements of empty sample holders. The ratio of capture signal-to-background is an important quantity due to the fact that there is uncertainty in the background measurement since the empty can does not account for interactions of off-energy background neutrons with the sample. The signal-to-background ratios for these measurements are shown in the lower plot of Figure 1 and Figure 2 for thermal and epithermal measurements, respectively.

    b. Transmission data

Raw transmission data were corrected for dead time, normalized to the beam monitors, summed over all cycles, background subtracted, and processed into transmission as given by Eq. (2).

$$T_i = \frac{\left(C_i^S - K_S B_i - B_S\right)}{\left(C_i^O - K_O B_i - B_O\right)}, \quad (2)$$

where
$T_i$, the transmission in time-of-flight channel $i$,
$C_i^S$ and $C_i^O$ are the dead-time corrected and monitor-normalized counting rates of the sample and open measurements in channel $i$, respectively,
$B_i$ is the unnormalized, time-dependent background counting rate in channel $i$,
$B_S$ and $B_O$ are the steady state background counting rates for sample and open measurements, respectively, and
$K_S$ and $K_O$ are the normalization factors for the sample and open background measurements.

Background determination and subtraction is more difficult to determine in transmission than capture, where the background, before subtraction, is visible between resonances. The relative effect of the uncertainty in the background measurement is given by the signal-to-background ratio. These ratios are provided for thermal measurements in Figure 1 and for epithermal measurements in Figure 2. The upper plot in each figure shows the component counting rates. The structure in the plot of the open beam counting rate is due to the fixed notch, see Table 3.

### 1. Transmission Background

The background in the transmission measurements was measured explicitly at the black resonance of a 'fixed' notch, a filter placed in the beam throughout the Cd measurements for the express purpose of providing a normalization point for the background shape. The energy dependent background shape was determined by a separate series of experiments that consisted of a package of black resonance filters placed in the neutron beam. The notch thickness and black resonances for thermal measurements included 0.91-mm of cadmium at 0.178 eV, 0.25-mm of indium at 1.4 eV, 0.18-mm of silver at 5.2 eV, and 0.13-mm of tungsten at 18.8 eV. The notch thicknesses and black resonances for epithermal measurements included 0.18-mm silver at 5.2 eV, 0.13-mil tungsten at 18.8 eV, 0.25-mil cobalt at 130 eV, and 1.0-mm of an 80 weight % Mn, 20 weight % Cu alloy at 336 eV. Single- and double-thicknesses of these notch filters were placed in the beam and measured with each Cd sample as well as open beam. The one-notch and two-notch data were used to extrapolate to zero-notch thickness.[12] The resulting background shape was normalized to the fixed notch given in the rightmost column of Table 3.

### IV. Results
#### a. Resonance Parameters

The SAMMY 8.0[13] Bayesian multi-level R-matrix code was used to extract resonance parameters from the neutron capture and transmission data. The analysis employed the experimental resolution, Doppler broadening, self-shielding, multiple scattering, and propagated uncertainty parameter features of SAMMY. The resulting resonance parameters of cadmium are listed in Table 5. The first four columns are resonance energies and uncertainties, the next four columns are radiation widths and uncertainties, then four columns of neutron width information, followed by isotope, spin and parity ($J^\pi$), and angular momentum ($\ell$). The measured value, the Bayesian and external uncertainties (see Section IV.c), and the ENDF/B-VII.1 values[14] are given in Table 5 for each of the resonance parameters, resonance energy, neutron width, $\Gamma_n$, and radiation width, $\Gamma_\gamma$. The external uncertainties described in Section IV.c represent the sample-to-sample consistency within the current data.

A graphical overview of the data, the SAMMY fit representing the NNL/RPI resonance parameters, and those from ENDF/B-VII.1 in the thermal region are shown in Figure **3**. The 2200 m/s cross section is defined at 0.0253 eV. Its value is dictated by the fit to the strong resonance in Cd113 at 0.178 eV. The value could also be influenced by the presence of negative energy resonances, also known as bound levels. There are no negative energy resonances in Cd113 in any of the libraries listed in Table 2. The best fit to the 0.178 eV resonance data slightly underestimates the transmission for the 0.10 mm Cd sample at 0.0253 eV, see Figure **3**. . It can be seen from Figure **3** that the transmission of this sample at 0.0253 eV is approximately 0.3. This sample is one of the most sensitive since it is thick enough to show a deep resonance in transmission without as much reliance on the background fit as thicker samples with transmissions near zero.[15] No negative energy resonance was added for Cd113 because an additional resonance would only move the calculated transmission lower and not improve the fit. The contributions from the other resonances to the thermal cross section of natural Cd is <<1%. Therefore, none of the negative energy resonances were fitted. The JENDL-4.0[16] negative energy resonances were adopted because they gave the best fit to all of the data among the libraries listed in Table 2.

An overview of the epithermal data, the SAMMY fit representing the NNL/RPI resonance parameters, and those from ENDF/B-VII.1 in the 0-600 eV region is shown in Figure 4. The 392 eV resonance is very strong, but it is only about 6% capture. A closer look at the four resonances between 83 and 93 eV is shown in Figure 5. The strong resonance at ~90 eV is mostly scattering. The shape of the capture data for the thickest sample is distorted due to multiple scattering.

The region from 600-1000 eV is shown in Figure 6. These transmission data were analyzed exclusively to produce the resonance parameters given in Table 5. Capture data were only included in the analysis below 600 eV in order to ensure that <2% of scattering neutrons penetrated the boron carbide liner and were captured in the NaI and erroneously counted as Cd capture events.[17] Reference 17 describes MCNP modeling supported by measurements with a Pb scatterer inside the NaI capture detector.

The resonance parameter fit with the SAMMY code included a description of the experimental resolution and an uncertainty propagation from eight specific sources. The resolution function differed for capture, thermal transmission, and epithermal transmission measurements. The sources of experimental uncertainty were resolution function, normalization, background, flight path length, zero time, burst width, sample thickness, and effective temperature.

Although their uncertainties were propagated none of the values of these eight sources of experimental uncertainty were fitted with the exception of capture normalization in the thermal region. The normalization for the data from all of the samples in a specific capture experiment were varied together to get the best fit in combination with the unnormalized transmission data. The resulting adjustments to the thermal capture were 0.2% for week 1 and 0.5% for week 2.

The effective temperature used in the analysis was 298.9 K. The experiment resolution functions represent the effects of the LINAC electron pulse width, the emission time in the moderator, the TOF channel width, and any broadening by the detector system. The resolution function for thermal and epithermal capture, as well as epithermal transmission was characterized in SAMMY as a Gaussian distribution in time plus an exponential tail. The widths of the exponential tails of the distributions were fitted to measurements of depleted uranium.[11,18] In the thermal transmission measurement the photomultiplier tube was in the neutron beam and was represented by a more complex resolution function.[19] A functional representation of the RPI resolution function is included in the SAMMY code and was used to analyze the thermal transmission data.

b. Radiation Width Determination

Radiation width information was extracted from the data wherever the sensitivity criterion of $\Gamma_\gamma / \Gamma_n < 5$ was satisfied or whenever the resonance energy was low enough that the resolution width was small and the radiation width was derived directly from the observed width of the resonance.[2,11] The fitted values for these radiation widths and their uncertainties are given in Table 5. Whenever no uncertainty is shown in Table 5 the criterion was not met, $\Gamma_\gamma$ was not varied, and the value from ENDF/B-VII.1 was used. The sensitive $\Gamma_\gamma$s were used to determine average $\Gamma_\gamma$s for each isotope and spin state with three or more resonances in Table 6. The uncertainties in the average radiation widths given in Table 6 are the standard deviations of each distribution rather than the standard error on the mean because they are intended to represent the uncertainty in the expected width of the next identified insensitive resonance from each population of a particular isotope and spin state. Table 6 also gives the number of degrees of freedom, DF, in each chi-squared distribution given by two divided by the variance in the distribution of $\Gamma_\gamma$, 2/var($\Gamma_\gamma$). The total number of resonances in the measured energy range, 0-1 keV, and the number of resonances sensitive to $\Gamma_\gamma$ are also given in Table 6 for each isotope and spin.

c. External uncertainties; the weighted standard deviation of individual sample fits uncertainty method

The external uncertainty, $\Delta X_{ext}$, of a resonance parameter, $X$, is given by Eq. (3),

$$\Delta X_{ext} = \sqrt{\frac{\sum_1^n (X_i - <X>)^2}{(\Delta X_i)^2} / (n-1) \sum_1^n \frac{1}{(\Delta X_i)^2}}, \quad (3)$$

where

$$<X> = \frac{\sum_1^n X_i}{(\Delta X_i)^2} / \sum_1^n \frac{1}{(\Delta X_i)^2},$$

and $<X>$ is the weighted mean value of $X$, $X_i$ is the value of $X$ for sample $i$, $\Delta X_i$ is the SAMMY uncertainty in $X$ for sample $i$, and $n$ is the number of samples. The value of each resonance parameter was fitted to each sample individually, and the resulting distribution of values for each isotope was analyzed. The resulting external uncertainties are given in square brackets in Table 5.

Inspection of Table 5 shows that the external uncertainties are generally larger than the Bayesian uncertainties for larger resonances at low energies with many samples and many points defining the resonance. However, above 100 eV the Bayesian uncertainties from SAMMY are usually larger.

These are weighted standard deviations, not formal errors whose magnitudes can be used to imply confidence intervals. The external uncertainties in Eq. (3) are weighted by the inverse variances from the Bayesian fit whose values were propagated from all of the sources of uncertainty discussed in Section IV.a. However, these variances were not sufficient, in the Bayesian analysis, to account for the deviations among samples. The Bayesian method is characterized by the use of random variables to model all sources of uncertainty. Therefore, much of the uncertainty that is quantified by the weighted standard deviation method could be due to systematic uncertainties.

d. Capture Resonance Integrals and Thermal Total Cross Section

Infinitely dilute capture resonance integrals (RIs) have been calculated from Eq. (4),

$$RI = \int_{0.5eV}^{20MeV} \sigma_C(E) \frac{dE}{E} \quad ,\qquad(4)$$

where $\sigma_C(E)$ is the capture cross section in barns calculated from the derived resonance parameters, Doppler broadened to 300K, and $E$ is energy in eV. The cross section was calculated from the resonance parameters shown in Table 5. The results for the capture resonance integral are given in Table 7. The current results shown in the rightmost column of Table 7 reflect ENDF/B-VII.1 resonance parameters above the maximum measured resonance energy of 1 keV. For Cd113 the current measurement agrees with the current ENDF value for capture resonance integral and is closer to JENDL-4.0 and the Atlas[20] than it is to JEFF-3.2.[21] For Cd111 the current results agree with the recent ENDF and JEFF libraries and not with JENDL or the Atlas.

The thermal total cross section results from the current measurement are given in Table 8 where they are compared to the values from the nuclear data libraries. The thermal cross section of Cd113 overwhelms the contributions from all other isotopes. The current measurements give a thermal cross section for Cd113 that is within the measured uncertainty of the JENDL-4.0 value and within 1% of the ENDF/B-VII.1 and JEFF-3.2 values.

Both resonance integrals and thermal cross sections were calculated by processing the resonance parameters into Doppler-broadened cross sections using the NJOY[22] code. Integration and interpolation were accomplished using the INTER[23] code. The uncertainty in each value in Table 7 and Table 8 was determined using a Monte Carlo sampling method where the NJOY and INTER codes were run repeatedly while sampling resonance parameters within the larger of their Bayesian or external uncertainties from Table 5.

e. Nuclear Level Statistics: Level Density and Strength Functions

The level density "staircase plots" for the Cd isotopes of mass number 110-114 are given in Figure 7. The average level spacing, $D_0$, is the measured energy range divided by the number of resonances. The average level spacing is represented in Figure 7 as the inverse slope of the solid line shown in each plot. The uncertainty band around the average is represented by dashed lines. The uncertainty in the average level spacing, $\Delta D_0$, was determined using Eq. (5) by assuming a Wigner distribution,

$$\Delta D_0 = \frac{0.52 D_0}{\sqrt{N}}, \qquad (5)$$

where $N$ is the number of levels.[20] Figure 7 shows that the JENDL-4.0 evaluated library has many more resonances in isotopes Cd110, Cd111, Cd112 and Cd114 beyond the measured energy range than were adopted in JEFF-3.2 or ENDF/B-VII.1, which was used as a starting point for the current analysis. These additional levels were also present in some isotopes for the earlier versions of ENDF.

The neutron strength function, $S_0$, is defined as the average reduced neutron width, $<\Gamma_n^0>$, divided by the average level spacing, as shown in Eq. (6).

$$S_0 = \frac{<\Gamma_n^0>}{D_0}, \qquad (6)$$

where the reduced neutron width, $\Gamma_n^0$, is the neutron width of a resonance divided by the square root of the resonance energy. The strength functions for the Cd isotopes are given in Table 10. The strength function was computed from the slope of the cumulative distribution of reduced neutron widths as described in Appendix A, Section A.2. The values of the strength function given in Table 10 for the current NNL/RPI measurement agree within uncertainties with all libraries dating back at least to the release of ENDF/B-VI.8 in 2001.

The uncertainties in Table 10 are large because the slope is fit to a constant but appears to vary with energy. Therefore, the strength function uncertainties (see Appendix A, Section A.2) account for the nonlinearity of the distribution and reflect the variations in strength functions calculated over various, smaller energy ranges.

The maximum resonance energy used for each library in the calculation of the strength functions in Table 10 is given in Table 11. The values in Table 11 define a region of linearity that was determined for each isotope and library by inspection of Figure 7 and represent the approximate energies where each evaluation begins to be missing levels for each isotope. The number of resonances in the region of linearity for Cd110-113 is given in Table 12. The large uncertainties in the values of the slopes of lines fitting the data from Cd114, the lower plot in Figure 7, render the strength functions for Cd114 isotopes inconclusive.

Only one resonance, at 387 eV, has been added to those listed in ENDF/B-VII.1. So, the concern over nonlinearity of level spacing is not limited to the present measurement. It is a characteristic of all of the libraries shown in Figure 7. The Atlas values in Table 10 are taken directly from the references without any specification of energy region. However, some insight was gathered about the energy region corresponding to the determination of the Atlas strength functions because the ENDF/B-VI.8 evaluation[24] is based in large part upon Reference 25, and the ENDF/B-VII.0 evaluation[26] is based in large part upon Reference 20. For Cd113, analysis of the resonances from ENDF/B-VI.8 indicate that the strength function cited from Reference 25 is consistent with an analysis up to 2.3 keV. Furthermore, also for Cd113, analysis of the resonances from ENDF/B-VII.0 indicate that the strength function cited from Reference 20 is consistent with an analysis up to approximately 1 keV.

The adherence of the reduced neutron width distributions to the theory of Porter and Thomas[27] is shown in Figure 8 for each isotope. For the two isotopes with a significant number of resonances in the measured range Cd111 shows a reasonably good fit to the Porter Thomas theory of a chi-squared distribution with one degree of freedom, but Cd113 appears to contain several more small amplitude resonances and several less large amplitude resonances than fit the Porter Thomas compound nucleus theory. The maximum energy of the resonances included in Figure 8 for each isotope and each library is given in Table 11, based on the limit of linear behavior of the staircase plot in Figure 7. The points in Figure 8 were normalized such that every library appears to have the same total number of resonances. The number of resonances in each plot is given in Table 12.

The only resonance that was identified in addition to those in the ENDF/B-VII.1 database occurred at 387 eV and was assigned as a p-wave in Cd113 with a $J^\pi$ value of $2^-$.

    f.   Nuclear Radius fits

Nuclear radii for all isotopes were fitted from epithermal transmission data in the regions between resonances. The results for p-wave resonances were unchanged from ENDF/B-VII.1. For s-wave resonances results are shown in Table 9. Changes less than 1% were observed for most isotopes relative to ENDF/B-VII.1. The most significant changes to the nuclear radii were for Cd112, Cd113, and Cd114, all reduced by about 5%. Cd111 showed a reduction of 3% in nuclear radius from the initial ENDF/B-VII.1 values. No external R-function was applied.

## V.   Conclusions

The current measurements of natural cadmium complement a long series of measurements and evaluations that are forming a consensus on the resonance parameters of the Cd isotopes in general and the 0.178 eV resonance in Cd113 in particular. Results include resonance parameters, staircase plots of level densities, strength functions, reduced neutron width distribution comparisons to Porter Thomas, average radiation widths for particular isotopes and spins, capture resonance integrals, and thermal total cross sections.

For Cd113 the capture resonance integral from the current measurement agrees with the ENDF/B-VII.1 value and is closer to JENDL-4.0 and the Atlas[20] than it is to JEFF-3.2. For Cd111 the current result agrees with the recent ENDF and JEFF libraries and not with JENDL or the Atlas.

The thermal cross section of Cd113 overwhelms the contributions from all other isotopes of natural Cd. The current measurements give a thermal cross section for Cd113 that is within the measured uncertainty of the JENDL-4.0 value and within 1% of the ENDF/B-VII.1 and JEFF-3.2 values.

**Table 1- A review of historical measurements for the 0.178 eV resonance in Cd113 in reverse chronological order. The present results are bolded.**

| Source/ authors | date | Refer-ence no. | Energy, eV | $\Gamma_\gamma$, meV | $\Gamma_n$, meV |
|---|---|---|---|---|---|
| **NNL/RPI** | **2017** | **5** | **0.1779 ± 0.0002** | **112.4 ± 0.4** | **0.638 ± 0.008** |
| Kopecky et al. | 2009 | 28 | 0.1787 ± 0.0001 | 113.5 | 0.640 ± 0.004 |
| Harz and Priesmeyer | 1974 | 29 | 0.1783 ± 0.0002 | 112.9 | 0.650 ± 0.005 |
| Widder and Brunner | 1968 | 30 | 0.1776 ± 0.0006 | 113.7 | 0.618 ± 0.003 |
| Akyüz et al. | 1967 | 31 | 0.181 ± 0.003 | 108.7 ± 0.3 | 0.791 ± 0.032 |
| Shchepkin et al. | 1966 | 32 | 0.178 ± 0.002 | 113 ± 5 | 0.65 ± 0.02 |
| Meservey | 1954 | 33 | 0.177 ± 0.005 | 109.4 | 0.608 |
| Brockhouse | 1953 | 34 | 0.180 ± 0.003 | 112.4 | 0.648 |
| Rainwater et al. | 1947 | 35 | 0.176 ± 0.002 | 114.4 | 0.599 |
| Rainwater et al. | 1946 | 36 | 0.180 ± 0.005 | 111.4 | 0.646 |

**Table 2- Review of evaluated values for the 0.178 eV resonance in Cd113 in reverse chronological order. The present results are bolded. The NNL/RPI neutron width agrees with JENDL-4.0 and JEFF-3.2. The NNL/RPI resonance energy agrees with the older ENDFs.**

| Source | Energy, eV | $\Gamma_\gamma$, meV | $\Gamma_n$, meV | Date |
|---|---|---|---|---|
| **NNL/RPI** | **0.1779 ± 0.0002** | **112.4 ± 0.5** | **0.638 ± 0.008** | **2017** |
| JEFF-3.2 | 0.1787 | 113.5 | 0.64 | 2014 |
| ENDF/B-VII.1 | 0.1787 | 113.5 | 0.63362 | 2011 |
| JENDL-4.0 | 0.1787 | 113.5 | 0.64 | 2011 |
| ENDF/B-VII.0 | 0.178 | 113 | 0.65333 | 2006 |
| ENDF/B-VI.8 | 0.178 | 113.65 | 0.65333 | 2001 |

**Table 3- Cadmium Experimental Details**

| Experiment | Overlap Filter | Neutron-Producing Target | Electron Pulse Width (ns) | Ave. Beam Current (µA) | Beam Energy (MeV) | Pulse Repetition Rate (pulses/s) | Flight Path Length (m) | Normalization Point for Transmission Background or Capture Flux (eV) |
|---|---|---|---|---|---|---|---|---|
| Epithermal Transmission | Boron Carbide | Bare Bounce | 53 ± 3 | 22 | 56 | 225 | 25.589 ± 0.007 | 45 eV resonance in Mo95 fixed notch |
| Thermal Transmission Week 1 of 3 | None | Enhanced Thermal Target | 1900 ± 100 | 12 | 50 | 25 | 14.973 ± 0.004 | 18.8 eV resonance in W186 |
| Thermal Transmission Week 2 of 3 | None | Enhanced Thermal Target | 920 ± 50 | 6 | 55 | 25 | 14.973 ± 0.004 | 5.2 eV resonance in Ag109 |
| Thermal Transmission Week 3 of 3 | None | Enhanced Thermal Target | 900 ± 70 | 6 | 60 | 25 | 14.973 ± 0.004 | 5.2 eV resonance in Ag109 |
| Epithermal Capture Week 1 of 2 | Boron Carbide | Bare Bounce | 49 ± 2 | 23 | 52 | 225 | 25.564 ± 0.006 | 27 eV resonance in Cd111 in the 200-mil sample |
| Epithermal Capture Week 2 of 2 | Boron Carbide | Bare Bounce | 48.3 ± 0.1 | 23 | 54 | 225 | 25.564 ± 0.006 | 27 eV resonance in Cd111 in the 200-mil sample |
| Thermal Capture Week 1 of 2 | None | Enhanced Thermal Target | 1400 ± 100 | 11 | 50 | 25 | 25.444 ± 0.004 | 0.178 eV resonance in Cd113 in 20-mil sample data |
| Thermal Capture Week 2 of 2 | None | Enhanced Thermal Target | 770 ± 40 | 8.5 | 54 | 25 | 25.444 ± 0.004 | 0.178 eV resonance in Cd113 in 20-mil sample data |

**Table 4- Sample Details**

| Nominal thickness (mm) | Areal density (atoms/barn) | Measurements |
|---|---|---|
| 0.0254 (0.001 in.) | $1.133 \times 10^{-4} \pm 1 \times 10^{-7}$ | Thermal capture |
| 0.0508 (0.002 in.) | $2.329 \times 10^{-4} \pm 2 \times 10^{-7}$ | Thermal transmission and capture |
| 0.1016 (0.004 in.) | $4.280 \times 10^{-4} \pm 4 \times 10^{-7}$ | Thermal transmission and capture, and epithermal capture |
| 0.254 (0.010 in.) | $1.180 \times 10^{-3} \pm 1 \times 10^{-6}$ | Thermal transmission and epithermal capture |
| 0.508 (0.020 in.) | $2.393 \times 10^{-3} \pm 2 \times 10^{-6}$ | Thermal and epithermal transmission, Thermal and epithermal capture |
| 1.27 (0.050 in.) | $5.853 \times 10^{-3} \pm 6 \times 10^{-6}$ | Epithermal capture |
| 2.54 (0.100 in.) | $1.191 \times 10^{-2} \pm 1 \times 10^{-5}$ | Epithermal transmission |
| 5.08 (0.200 in.) | $2.430 \times 10^{-2} \pm 3 \times 10^{-5}$ | Epithermal transmission and capture |
| 10.16 (0.400 in.) | $4.426 \times 10^{-2} \pm 5 \times 10^{-5}$ | Epithermal transmission |

**Table 5– Resonance parameters for cadmium compared with ENDF/B-VII.1 parameters. Two uncertainties are given for each parameter, the Bayesian uncertainty from the SAMMY fit and an external error, in brackets, which conveys the agreement among individual sample fits.**

| E, eV | ΔE Bayesian [external] | Eendf | $\Gamma_\gamma$ | Δ $\Gamma_\gamma$ Bayesian [external] | $\Gamma_\gamma$endf | $\Gamma_n$ | Δ $\Gamma_n$ Bayesian [external] | $\Gamma_n$endf | isotope | $J^\pi$ | $\ell$ |
|---|---|---|---|---|---|---|---|---|---|---|---|
| -700 | | -11.7800 | 110 | | 112 | 700 | | 0.4485 | 108 | 0.5 | 0 |
| -550 | | -195.8600 | 37 | | 50.34 | 3000 | | 177.0000 | 116 | 0.5 | 0 |
| -225 | | -400.5400 | 55 | | 50 | 100 | | 2000.0000 | 114 | 0.5 | 0 |
| -125 | | -12.1640 | 100 | | 76 | 870 | | 3.4390 | 112 | 0.5 | 0 |
| -100 | | -97.4250 | 180 | | 160 | 120 | | 129.7000 | 106 | 0.5 | 0 |
| -20 | | -9.5698 | 110 | | 73 | 40 | | 9.6580 | 110 | 0.5 | 0 |
| -4 | | -27.9430 | 68 | | 129 | 1 | | 187.3000 | 111 | 1.0* | 0 |
| 0.1779 | 0.00002 [0.0002] | 0.1787 | 112.4 | 0.03 [ 0.4] | 113.5 | 0.6379 | 0.0003[ 0.008] | 0.6336 | 113 | 1.0 | 0 |
| 4.56 | 0.09 | 4.5630 | 163 | | 163 | 0.0014 | 0.0009 | 0.0014 | 111 | -1.0 | 1 |
| 6.95 | 0.06 | 6.9546 | | | 143 | | | 0.0009 | 111 | -2.0 | 1 |
| 7.0 | 0.1 | 7.000 | | | 160 | | | 0.0002 | 113 | -1.0 | 1 |
| 18.362 | 0.002 [ 0.002] | 18.365 | 95 | | 95 | 0.181 | 0.002[ 0.004] | 0.1880 | 113 | 1.0 | 0 |
| 21.87 | 0.03 [ 0.02] | 21.831 | 82 | | 82 | 0.0054 | 0.0005[0.0001] | 0.0057 | 113 | -2.0 | 1 |
| 27.5567 | 0.0005 [0.0007] | 27.570 | 131 | | 131.1 | 4.88 | 0.01[ 0.07] | 4.9120 | 111 | 1.0 | 0 |
| 28.97 | 0.01 [ 0.004] | 28.979 | 70 | | 70 | 0.030 | 0.001[0.0001] | 0.0269 | 116 | -1.5 | 1 |
| 43.4 | 0.1 [ 0.009] | 43.380 | 128 | | 128 | 0.0039 | 0.0004[0.0001] | 0.0038 | 113 | -2.0 | 1 |
| 49.76 | 0.07 [ 0.01] | 49.770 | 160 | | 160 | 0.064 | 0.005[0.0005] | 0.0600 | 113 | 0.0 | 1 |
| 54.24 | 0.05 [ 0.01] | 54.200 | 115 | | 115 | 0.18 | 0.01[0.0008] | 0.1575 | 108 | -1.5 | 1 |
| 56.10 | 0.05 [ 0.02] | 56.230 | 169 | | 169 | 0.055 | 0.005[0.0002] | 0.0537 | 113 | -1.0 | 1 |
| 56.353 | 0.009 [ 0.005] | 56.425 | 72 | | 71.5 | 0.033 | 0.001[0.0008] | 0.0375 | 114 | -1.5 | 1 |
| 63.688 | 0.001 [ 0.002] | 63.700 | 90 | | 90 | 3.22 | 0.02[ 0.04] | 3.4667 | 113 | 1.0 | 0 |
| 66.759 | 0.0008 [ 0.001] | 66.780 | 63 | | 62.8 | 8.04 | 0.03[ 0.07] | 8.4760 | 112 | 0.5 | 0 |
| 69.56 | 0.01 [ 0.01] | 69.590 | 130 | | 130 | 0.170 | 0.007[ 0.004] | 0.1760 | 111 | 1.0 | 0 |
| 81.5 | 0.2 [ 0.004] | 81.520 | 113 | | 113 | 0.0045 | 0.0004[0.0001] | 0.0043 | 113 | -2.0 | 1 |
| 82.42 | 0.02 [ 0.008] | 82.466 | 90 | | 90.5 | 0.039 | 0.002[0.0006] | 0.0409 | 112 | -1.5 | 1 |
| 83.289 | 0.005 [ 0.004] | 83.315 | 90 | | 90.5 | 0.176 | 0.003[ 0.003] | 0.1870 | 112 | -1.5 | 1 |
| 84.855 | 0.001 [ 0.002] | 84.910 | 144 | 2 [ 2 ] | 110 | 31.3 | 0.1[ 0.5] | 29.8670 | 113 | 1.0 | 0 |
| 86.165 | 0.002 [ 0.003] | 86.185 | 130 | | 130.1 | 3.10 | 0.02[ 0.02] | 3.3540 | 111 | 1.0 | 0 |
| 89.515 | 0.001 [ 0.002] | 89.549 | 75 | 0.4 [ 1 ] | 73.18 | 159 | 0.4[ 1 ] | 162.7000 | 110 | 0.5 | 0 |
| 98.3 | 0.1 [ 0.03] | 98.520 | 225 | | 225 | 0.044 | 0.003[0.0001] | 0.0336 | 113 | -2.0 | 1 |
| 99.450 | 0.001 [ 0.002] | 99.491 | 130 | | 129.6 | 14.4 | 0.06[ 0.2] | 14.5700 | 111 | 1.0 | 0 |
| 102.2 | 0.1 [ 0.03] | 102.30 | 117 | | 117 | 0.039 | 0.003[0.0003] | 0.0296 | 113 | -2.0 | 1 |

\* The ENDF/B-VII.1 $J^\pi$ for this resonance is 0.0

**Table 5 (cont.)– Resonance parameters for cadmium compared with ENDF/B-VII.1 parameters. Two uncertainties are given for each parameter, the Bayesian uncertainty from the SAMMY fit and an external error, in brackets, which conveys the agreement among individual sample fits.**

| E, eV | ΔE Bayesian [external] | Eendf | $\Gamma_\gamma$ | $\Delta \Gamma_\gamma$ Bayesian [external] | $\Gamma_\gamma$endf | $\Gamma_n$ | $\Delta \Gamma_n$ Bayesian [external] | $\Gamma_n$endf | $J^\pi$ isotope | $\ell$ |
|---|---|---|---|---|---|---|---|---|---|---|
| 102.963 | 0.006 [ 0.002] | 103.08 | 130 | | 130 | 1.13 | 0.02 [ 0.02] | 1.0670 | 111  1.0 | 0 |
| 106.6 | 0.1 [ 0.04] | 106.56 | 151 | | 151 | 0.030 | 0.002 [0.0001] | 0.0240 | 113  -2.0 | 1 |
| 108.304 | 0.002 [ 0.002] | 108.33 | 90 | | 90 | 10.8 | 0.05 [ 0.1] | 11.3600 | 113  1.0 | 0 |
| 114.91 | 0.03 [ 0.02] | 114.83 | 165 | | 165 | 0.151 | 0.007 [ 0.001] | 0.1120 | 111  -2.0 | 1 |
| 120.095 | 0.001 [ 0.002] | 120.13 | 49.9 | 0.3 [ 0.8] | 49.98 | 61.7 | 0.2 [ 0.8] | 60.8100 | 114  0.5 | 0 |
| 138.121 | 0.002 [ 0.004] | 138.17 | 130 | | 130 | 9.5 | 0.05 [ 0.1] | 9.5000 | 111  1.0 | 0 |
| 140.63 | 0.04 [ 0.02] | 140.87 | 165 | | 165 | 0.157 | 0.009 [ 0.002] | 0.1520 | 111  -2.0 | 1 |
| 143.063 | 0.004 [ 0.004] | 143.07 | 98 | | 98 | 9.2 | 0.1 [ 0.09] | 9.3600 | 113  0.0 | 0 |
| 158.743 | 0.003 [ 0.003] | 158.76 | 90 | | 90 | 8.56 | 0.06 [ 0.08] | 8.6267 | 113  1.0 | 0 |
| 164.187 | 0.002 [ 0.003] | 164.24 | 165 | 2 [ 2 ] | 127 | 61.5 | 0.5 [ 0.6] | 68.5800 | 111  1.0 | 0 |
| 166.7 | 0.3 [ 0.009] | 166.60 | 160 | | 160 | 0.016 | 0.002 [0.0001] | 0.0160 | 113  -2.0 | 1 |
| 192.820 | 0.002 [ 0.003] | 192.85 | 110 | 1 [ 2 ] | 110 | 182 | 1 [ 1 ] | 183.2000 | 113  0.0 | 0 |
| 196.2 | 0.1 [ 0.02] | 196.15 | 160 | | 160 | 0.13 | 0.01 [0.0009] | 0.1333 | 113  -1.0 | 1 |
| 203.1 | 0.3 [ 0.006] | 203.51 | 160 | | 160 | 0.053 | 0.005 [0.0003] | 0.0536 | 113  -2.0 | 1 |
| 203.74 | 0.07 [ 0.03] | 203.66 | 165 | | 165 | 0.17 | 0.01 [ 0.003] | 0.1840 | 111  -2.0 | 1 |
| 208.85 | 0.07 [ 0.04] | 208.71 | 165 | | 165 | 0.21 | 0.01 [ 0.002] | 0.2080 | 111  -2.0 | 1 |
| 211.8 | 0.2 [ 0.01] | 211.88 | 160 | | 160 | 0.11 | 0.01 [0.0001] | 0.1040 | 113  -1.0 | 1 |
| 215.206 | 0.003 [ 0.004] | 215.23 | 177 | 4 [ 1 ] | 110 | 26.6 | 0.2 [ 0.3] | 26.9330 | 113  1.0 | 0 |
| 225.078 | 0.003 [ 0.005] | 225.15 | 242 | 5 [ 5 ] | 130.5 | 32.1 | 0.2 [ 0.4] | 32.1000 | 111  1.0 | 0 |
| 226.434 | 0.005 [ 0.007] | 226.51 | 53 | 2 [ 0.5] | 55.9 | 18.9 | 0.3 [ 0.2] | 20.5700 | 112  0.5 | 0 |
| 226.95 | 0.02 [ 0.008] | 226.90 | 72 | | 71.5 | 1.28 | 0.08 [ 0.02] | 1.5820 | 114  -1.5 | 1 |
| 230.984 | 0.006 [ 0.005] | 231.03 | 70 | | 70 | 6.53 | 0.07 [ 0.05] | 6.6860 | 110  0.5 | 0 |
| 232.23 | 0.04 [ 0.04] | 232.41 | 98 | | 98 | 4.8 | 0.3 [ 0.04] | 4.2400 | 113  0.0 | 0 |
| 233.399 | 0.005 [ 0.009] | 233.47 | 167 | 7 [ 1 ] | 130 | 47 | 1 [ 0.9] | 51.3000 | 111  1.0 | 0 |
| 233.59 | 0.02 [ 0.02] | 233.60 | 170 | 10 [ 1 ] | 153 | 360 | 10 [ 6 ] | 349.0000 | 108  0.5 | 0 |
| 237.8 | 0.2 [ 0.04] | 237.87 | 160 | | 160 | 0.101 | 0.009 [0.0004] | 0.1000 | 113  -2.0 | 1 |
| 252.3 | 0.2 [ 0.05] | 252.68 | 160 | | 160 | 0.21 | 0.02 [0.0004] | 0.1867 | 113  -1.0 | 1 |
| 261.048 | 0.003 [ 0.004] | 261.07 | 177 | 4 [ 1 ] | 110 | 33.7 | 0.3 [ 0.3] | 35.0670 | 113  1.0 | 0 |
| 269.313 | 0.005 [ 0.005] | 269.35 | 109 | 3 [ 0.8] | 100 | 71.2 | 0.9 [ 0.6] | 70.0000 | 113  0.0 | 0 |
| 271.4 | 0.3 [ 0.008] | 271.50 | 160 | | 160 | 0.24 | 0.02 [0.0002] | 0.2080 | 113  -2.0 | 1 |
| 275.595 | 0.004 [ 0.005] | 275.69 | 136 | | 136.5 | 20.0 | 0.2 [ 0.2] | 20.4300 | 111  1.0 | 0 |
| 282.12 | 0.06 [ 0.04] | 281.83 | 98 | | 98 | 2.3 | 0.1 [ 0.03] | 1.9280 | 113  0.0 | 0 |
| 286.83 | 0.06 [ 0.05] | 286.66 | 165 | | 165 | 0.41 | 0.02 [ 0.005] | 0.3790 | 111  -2.0 | 1 |
| 289.4 | 0.3 [ 0.03] | 289.64 | 160 | | 160 | 0.051 | 0.005 [0.0001] | 0.0480 | 113  -2.0 | 1 |
| 291.60 | 0.009 [ 0.01] | 291.61 | 98 | | 98 | 6.06 | 0.09 [ 0.05] | 5.8667 | 113  1.0 | 0 |
| 311.52 | 0.009 [ 0.03] | 311.66 | 118 | 2 [ 3 ] | 145 | 509 | 6 [ 6 ] | 538.6000 | 108  0.5 | 0 |

**Table 5 (cont.)– Resonance parameters for cadmium compared with ENDF/B-VII.1 parameters. Two uncertainties are given for each parameter, the Bayesian uncertainty from the SAMMY fit and an external error, in brackets, which conveys the agreement among individual sample fits.**

| E, eV | ΔE Bayesian [external] | Eendf | $\Gamma_\gamma$ | $\Delta \Gamma_\gamma$ Bayesian [external] | $\Gamma_\gamma$endf | $\Gamma_n$ | $\Delta \Gamma_n$ Bayesian [external] | $\Gamma_n$endf | isotope | $J^\pi$ | $\ell$ |
|---|---|---|---|---|---|---|---|---|---|---|---|
| 313.0 | 0.1 [ 0.06] | 312.30 | 160 | | 160 | 0.90 | 0.07[ 0.002] | 0.6547 | 113 | -1.0 | 1 |
| 313.76 | 0.08 [ 0.04] | 313.46 | 165 | | 165 | 0.37 | 0.04[ 0.003] | 0.3870 | 111 | -2.0 | 1 |
| 331.99 | 0.01 [ 0.006] | 332.08 | 176 | | 176 | 25.0 | 0.4[ 0.3] | 23.3300 | 111 | 0.0 | 0 |
| 336.62 | 0.05 [ 0.03] | 336.88 | 130 | | 130 | 1.37 | 0.07[ 0.04] | 1.4680 | 111 | 1.0 | 0 |
| 337.7 | 0.3 [ 0.05] | 337.50 | 98 | | 98 | 1.1 | 0.1[ 0.003] | 1.0750 | 108 | -1.5 | 1 |
| 339.45 | 0.06 [ 0.04] | 339.86 | 79 | | 79 | 0.39 | 0.02[ 0.005] | 0.3594 | 110 | -1.5 | 1 |
| 343.6 | 0.2 [ 0.02] | 343.79 | 160 | | 160 | 0.14 | 0.01[0.0006] | 0.1360 | 113 | -2.0 | 1 |
| 351.7 | 0.5 [ 0.007] | 351.61 | 160 | | 160 | 0.029 | 0.003[0.0001] | 0.0288 | 113 | -2.0 | 1 |
| 356.115 | 0.005 [ 0.005] | 356.21 | 152 | 3 [ 0.7] | 120 | 48 | 0.7[ 2 ] | 48.1600 | 111 | 1.0 | 0 |
| 359.3 | 0.2 [ 0.06] | 359.32 | 160 | | 160 | 0.40 | 0.03[ 0.001] | 0.3733 | 113 | -1.0 | 1 |
| 369.546 | 0.006 [ 0.006] | 369.68 | 125 | 5 [ 0.3] | 89 | 18.3 | 0.3[ 0.3] | 19.6000 | 110 | 0.5 | 0 |
| 376.91 | 0.07 [ 0.03] | 376.82 | 98 | | 98 | 1.33 | 0.07[ 0.004] | 1.1067 | 113 | 1.0 | 0 |
| 384.4 | 0.4 [ 0.02] | 385.01 | 160 | | 160 | 0.13 | 0.01[0.0001] | 0.1187 | 113 | -1.0 | 1 |
| 387.85 | 0.08 [ 0.05] | NEW | 160 | | NEW | 1.19 | 0.06[ 0.004] | NEW | 113 | unassigned | 1 |
| 389.116 | 0.008 [ 0.007] | 389.17 | 165 | | 165 | 27.5 | 0.4[ 0.4] | 26.5500 | 111 | -1.0 | 1 |
| 392.376 | 0.006 [ 0.009] | 392.55 | 57.6 | 0.4 [ 0.6] | 55 | 965 | 2 [ 4 ] | 936.2000 | 114 | 0.5 | 0 |
| 410.34 | 0.04 [ 0.02] | 410.28 | 130 | | 130 | 2.4 | 0.1[ 0.01] | 2.2690 | 111 | 1.0 | 0 |
| 414.005 | 0.005 [ 0.009] | 414.05 | 105 | 1 [ 2 ] | 100 | 132 | 2 [ 0.8] | 126.0000 | 113 | 1.0 | 0 |
| 422.22 | 0.09 [ 0.03] | 422.73 | 160 | | 160 | 0.80 | 0.06[ 0.01] | 0.8000 | 113 | -2.0 | 1 |
| 423.3 | 0.1 [ 0.06] | 422.94 | 130 | | 130 | 0.93 | 0.07[ 0.009] | 0.9226 | 111 | 1.0 | 0 |
| 431.92 | 0.008 [ 0.01] | 432.01 | 151 | 7 [ 0.6] | 100 | 25.6 | 0.5[ 0.4] | 24.4000 | 113 | 1.0 | 0 |
| 438.29 | 0.02 [ 0.01] | 438.44 | 82 | | 82 | 9.4234 | 0.0002[0.0001] | 8.9330 | 111 | 1.0 | 0 |
| 442.915 | 0.006 [ 0.004] | 443.10 | 58.1 | 0.9 [ 0.3] | 62 | 63.3 | 0.9[ 0.5] | 61.3100 | 112 | 0.5 | 0 |
| 444.8 | 0.5 [ 0.009] | 443.23 | 165 | | 165 | 0.35 | 0.03[0.0001] | 0.2900 | 111 | -2.0 | 1 |
| 446.86 | 0.05 [ 0.04] | 447.11 | 98 | | 98 | 3.3 | 0.1[ 0.02] | 2.8000 | 113 | 1.0 | 0 |
| 452.66 | 0.02 [ 0.01] | 452.78 | 90 | | 90 | 1.52 | 0.03[ 0.02] | 1.7030 | 112 | -1.5 | 1 |
| 454.0 | 0.4 [ 0.03] | 453.21 | 165 | | 165 | 0.21 | 0.02[0.0001] | 0.1940 | 111 | -2.0 | 1 |
| 456.50 | 0.05 [ 0.02] | 456.66 | 134 | 9 [ 0.6] | 150 | 118 | 4 [ 2 ] | 102.0000 | 106 | 0.5 | 0 |
| 457.90 | 0.08 [ 0.05] | 457.75 | 98 | | 98 | 7.3 | 0.5[ 0.04] | 6.4800 | 113 | 0.0 | 0 |
| 466.4 | 0.2 [ 0.03] | 466.19 | 165 | | 165 | 0.52 | 0.04[ 0.003] | 0.4850 | 111 | -2.0 | 1 |
| 478.07 | 0.07 [ 0.04] | 478.22 | 130 | | 130 | 5.3 | 0.2[ 0.05] | 5.1670 | 111 | 1.0 | 0 |
| 483.14 | 0.09 [ 0.03] | 483.07 | 130 | | 130 | 2.0 | 0.1[ 0.02] | 1.7050 | 111 | 1.0 | 0 |
| 484.79 | 0.06 [ 0.04] | 484.93 | 130 | | 130 | 2.9 | 0.1[ 0.04] | 3.1050 | 111 | 1.0 | 0 |
| 489.9 | 0.2 [ 0.08] | 489.89 | 160 | | 160 | 1.16 | 0.08[ 0.003] | 0.9600 | 113 | -1.0 | 1 |
| 494.9 | 0.1 [ 0.07] | 495.08 | 98 | | 98 | 1.5 | 0.1[ 0.006] | 1.2133 | 113 | 1.0 | 0 |
| 501.4 | 0.5 [ 0.07] | 497.00 | 172 | | 172 | 1.1 | 0.1[ 0.001] | 1.0000 | 106 | -1.5 | 1 |

**Table 5 (cont.)– Resonance parameters for cadmium compared with ENDF/B-VII.1 parameters. Two uncertainties are given for each parameter, the Bayesian uncertainty from the SAMMY fit and an external error, in brackets, which conveys the agreement among individual sample fits.**

| E, eV | ΔE Bayesian [external] | Eendf | $\Gamma_\gamma$ | Δ $\Gamma_\gamma$ Bayesian [external] | $\Gamma_\gamma$endf | $\Gamma_n$ | Δ $\Gamma_n$ Bayesian [external] | $\Gamma_n$endf | isotope | $J^\pi$ | $\ell$ |
|---|---|---|---|---|---|---|---|---|---|---|---|
| 500.946 | 0.009 [ 0.007] | 501.00 | 133 | 5 [ 1 ] | 100 | 48 | 1 [ 0.8] | 49.2000 | 113 | 1.0 | 0 |
| 504.1 | 0.1 [ 0.09] | 503.64 | 79 | | 79 | 0.57 | 0.04[ 0.03] | 0.7179 | 110 | -1.5 | 1 |
| 518.5 | 0.2 [ 0.06] | 518.25 | 165 | | 165 | 0.79 | 0.06[ 0.007] | 0.6190 | 111 | -2.0 | 1 |
| 518.5 | 0.6 [ 0.02] | 518.30 | 160 | | 160 | 0.30 | 0.03[0.0007] | 0.2800 | 113 | -1.0 | 1 |
| 524.68 | 0.01 [ 0.01] | 524.80 | 150 | 8 [ 0.8] | 110 | 37.4 | 0.8[ 0.4] | 34.8000 | 113 | 1.0 | 0 |
| 528.0 | 0.2 [ 0.05] | 527.40 | 160 | | 160 | 1.19 | 0.09[ 0.003] | 0.9600 | 113 | -1.0 | 1 |
| 531.1 | 0.1 [ 0.05] | 530.99 | 165 | | 165 | 0.85 | 0.06[ 0.003] | 0.7740 | 111 | -2.0 | 1 |
| 533.5 | 0.2 [ 0.05] | 533.00 | 160 | | 160 | 3.5 | 0.3[ 0.009] | 3.0000 | 113 | 0.0 | 1 |
| 537.7 | 0.5 [ 0.05] | 537.50 | 160 | | 160 | 0.16 | 0.01[0.0002] | 0.1440 | 113 | -2.0 | 1 |
| 540.35 | 0.01 [ 0.008] | 540.46 | 175 | | 175 | 25.2 | 0.4[ 0.3] | 24.6800 | 111 | 1.0 | 0 |
| 543.0 | 0.6 [ 0.04] | 543.30 | 160 | | 160 | 0.25 | 0.02[0.0002] | 0.2400 | 113 | -1.0 | 1 |
| 543.2 | 0.2 [ 0.03] | 543.76 | 165 | | 165 | 0.86 | 0.06[ 0.004] | 0.7646 | 111 | -2.0 | 1 |
| 548.34 | 0.05 [ 0.05] | 548.57 | 130 | | 130 | 5.1 | 0.2[ 0.1] | 3.7470 | 111 | 1.0 | 0 |
| 551.797 | 0.009 [ 0.009] | 551.90 | 93 | 2 [ 1 ] | 110 | 133 | 2 [ 4 ] | 105.7300 | 113 | 1.0 | 0 |
| 553.7 | 0.7 [ 0.005] | 552.04 | 165 | | 165 | 0.10 | 0.01[0.0001] | 0.0967 | 111 | -2.0 | 1 |
| 561.7 | 0.7 [ 0.02] | 564.02 | 165 | | 165 | 0.10 | 0.01[0.0001] | 0.0967 | 111 | -2.0 | 1 |
| 565.67 | 0.04 [ 0.03] | 565.94 | 90 | | 90 | 1.21 | 0.04[ 0.02] | 1.2140 | 112 | -1.5 | 1 |
| 569.5 | 0.6 [ 0.01] | 568.01 | 165 | | 165 | 0.59 | 0.06[ 0.002] | 0.5820 | 111 | -2.0 | 1 |
| 567.78 | 0.09 [ 0.05] | 568.03 | 72 | | 72 | 0.97 | 0.04[ 0.02] | 1.0480 | 114 | -1.5 | 1 |
| 576.05 | 0.01 [ 0.02] | 576.18 | 260 | 10 [ 2 ] | 175 | 48.1 | 0.8[ 0.7] | 47.8400 | 111 | 1.0 | 0 |
| 578.3 | 0.5 [ 0.2] | 577.60 | 160 | | 160 | 0.32 | 0.03[0.0009] | 0.2960 | 113 | -2.0 | 1 |
| 583.2 | 0.4 [ 0.07] | 582.99 | 165 | | 165 | 0.42 | 0.04[0.0008] | 0.3870 | 111 | -2.0 | 1 |
| 584.8 | 0.4 [ 0.07] | 584.70 | 160 | | 160 | 0.68 | 0.06[0.0007] | 0.6267 | 113 | -1.0 | 1 |
| 591.8 | 0.2 [ 0.1] | 592.50 | 160 | | 160 | 0.76 | 0.06[0.0008] | 0.6800 | 113 | -2.0 | 1 |
| 598.7 | 0.1 [ 0.04] | 599.07 | 130 | | 130 | 7.3 | 0.5[ 0.05] | 6.7550 | 111 | 0.0 | 1 |
| 603.48 | 0.09 [ 0.01] | 603.83 | 130 | 10 [ 0.2] | 130 | 40 | 1 [ 1 ] | 34.5000 | 111 | 1.0 | 0 |
| 606.9 | 0.7 [ 0.007] | 606.95 | 165 | | 165 | 0.19 | 0.02[0.0002] | 0.1935 | 111 | -2.0 | 1 |
| 612.9 | 0.7 [ 0.03] | 613.00 | 160 | | 160 | 0.47 | 0.05[0.0007] | 0.4640 | 113 | -2.0 | 1 |
| 622.5 | 0.1 [ 0.02] | 622.77 | 140 | 10 [ 0.1] | 138 | 48 | 2 [ 1 ] | 47.8600 | 111 | 1.0 | 0 |
| 623.3 | 0.1 [ 0.01] | 623.70 | 100 | 10 [ 0.06] | 98 | 21 | 2 [ 0.2] | 20.5330 | 113 | 1.0 | 0 |
| 629.2 | 0.6 [ 0.06] | 629.90 | 98 | | 98 | 1.6 | 0.2[ 0.004] | 1.6000 | 113 | 1.0 | 0 |
| 634.74 | 0.09 [ 0.03] | 635.09 | 110 | 10 [ 0.3] | 110 | 810 | 20 [ 9 ] | 793.7000 | 106 | 0.5 | 0 |
| 652.6 | 0.3 [ 0.1] | 652.44 | 79 | | 79 | 1.9 | 0.2[ 0.004] | 1.9150 | 110 | -1.5 | 1 |
| 654.3 | 0.7 [ 0.1] | 653.86 | 165 | | 165 | 0.88 | 0.09[ 0.002] | 0.8740 | 111 | -2.0 | 1 |
| 661.3 | 0.7 [ 0.1] | 661.30 | 160 | | 160 | 4.2 | 0.4[ 0.02] | 4.1600 | 113 | 0.0 | 1 |
| 662.1 | 0.6 [ 0.1] | 662.90 | 98 | | 98 | 5.9 | 0.6[ 0.02] | 5.8400 | 113 | 0.0 | 0 |

**Table 5 (cont.)– Resonance parameters for cadmium compared with ENDF/B-VII.1 parameters. Two uncertainties are given for each parameter, the Bayesian uncertainty from the SAMMY fit and an external error, in brackets, which conveys the agreement among individual sample fits.**

| E, eV | ΔE Bayesian [external] | Eendf | $\Gamma_\gamma$ | $\Delta \Gamma_\gamma$ Bayesian [external] | $\Gamma_\gamma$endf | $\Gamma_n$ | $\Delta \Gamma_n$ Bayesian [external] | $\Gamma_n$endf | $J^\pi$ isotope | | $\ell$ |
|---|---|---|---|---|---|---|---|---|---|---|---|
| 670.8 | 0.8 [ 0.01] | 670.83 | 165 | | 165 | 1.3 | 0.1[0.0001] | 1.2660 | 111 | -2.0 | 1 |
| 670.92 | 0.09 [ 0.02] | 671.19 | 67 | 6 [ 0.5] | 59 | 439 | 5 [ 8 ] | 402.3000 | 114 | 0.5 | 0 |
| 676.2 | 0.3 [ 0.06] | 676.69 | 70 | | 70 | 11 | 1 [ 0.1] | 10.5300 | 116 | -1.5 | 1 |
| 678.0 | 0.4 [ 0.2] | 677.10 | 100 | 10 [ 0.009] | 98 | 20 | 2 [ 0.2] | 19.8670 | 113 | 1.0 | 0 |
| 685.8 | 0.8 [ 0.009] | 685.81 | 165 | | 165 | 0.10 | 0.01[0.0001] | 0.0967 | 111 | -2.0 | 1 |
| 687.5 | 0.8 [ 0.005] | 687.60 | 160 | | 160 | 0.24 | 0.02[0.0001] | 0.2400 | 113 | 0.0 | 1 |
| 689.0 | 0.6 [ 0.1] | 688.99 | 130 | | 130 | 3.5 | 0.3[ 0.02] | 3.5000 | 111 | 1.0 | 0 |
| 706.7 | 0.5 [ 0.2] | 706.68 | 130 | | 130 | 4.3 | 0.4[ 0.02] | 4.2900 | 111 | 1.0 | 0 |
| 709.1 | 0.8 [ 0.02] | 709.30 | 160 | | 160 | 0.27 | 0.03[0.0001] | 0.2720 | 113 | -2.0 | 1 |
| 718.19 | 0.004 [ 0.07] | 718.20 | 98 | | 98 | 14.4 | 0 [ 0.1] | 15.4670 | 113 | 1.0 | 0 |
| 723.4 | 0.2 [ 0.04] | 723.60 | 98 | | 98 | 19 | 1 [ 0.2] | 19.6000 | 113 | 1.0 | 0 |
| 734.0 | 0.2 [ 0.3] | 736.72 | 130 | | 130 | 11 | 2 [ 0.06] | 17.1400 | 111 | 1.0 | 0 |
| 737.3 | 0.1 [ 0.03] | 737.74 | 65 | 7 [ 0.3] | 67 | 332 | 5 [ 5 ] | 353.8000 | 112 | 0.5 | 0 |
| 740.0 | 0.8 [ 0.03] | 739.71 | 165 | | 165 | 0.88 | 0.09[0.0006] | 0.8740 | 111 | -2.0 | 1 |
| 752.0 | 0.1 [ 0.01] | 752.44 | 60 | 6 [ 0.8] | 54 | 400 | 5 [ 4 ] | 367.5000 | 114 | 0.5 | 0 |
| 754.1 | 0.9 [ 0.009] | 754.00 | 160 | | 160 | 0.27 | 0.03[0.0001] | 0.2720 | 113 | -2.0 | 1 |
| 758.6 | 0.6 [ 0.1] | 759.30 | 98 | | 98 | 14 | 1 [ 0.1] | 14.4400 | 113 | 0.0 | 0 |
| 761.7 | 0.3 [ 0.08] | 762.24 | 79 | | 79 | 5.3 | 0.5[ 0.04] | 5.5200 | 110 | -1.5 | 1 |
| 764.1 | 0.2 [ 0.1] | 764.83 | 121 | | 121 | 15 | 1 [ 0.4] | 15.8500 | 111 | 1.0 | 0 |
| 768.9 | 0.9 [ 0.01] | 768.90 | 160 | | 160 | 0.62 | 0.06[0.0008] | 0.6240 | 113 | -2.0 | 1 |
| 782.3 | 0.9 [ 0.02] | 782.20 | 160 | | 160 | 0.30 | 0.03[0.0001] | 0.3040 | 113 | -2.0 | 1 |
| 783.6 | 0.8 [ 0.1] | 783.21 | 165 | | 165 | 1.7 | 0.2[ 0.002] | 1.6810 | 111 | -2.0 | 1 |
| 790.1 | 0.6 [ 0.1] | 789.80 | 98 | | 98 | 19 | 2 [ 0.09] | 19.2000 | 113 | 1.0 | 0 |
| 790.3 | 0.2 [ 0.1] | 790.92 | 160 | 20 [ 0.1] | 161 | 52 | 3 [ 0.5] | 53.7000 | 111 | 1.0 | 0 |
| 799.7 | 0.1 [ 0.04] | 800.14 | 63 | 6 [ 0.2] | 69 | 453 | 7 [ 1 ] | 446.2000 | 110 | 0.5 | 0 |
| 809.3 | 0.3 [ 0.2] | 809.77 | 130 | 10 [ 0.06] | 130 | 42 | 3 [ 0.1] | 55.1600 | 111 | 1.0 | 0 |
| 809.9 | 0.7 [ 0.2] | 810.97 | 90 | | 90 | 3.8 | 0.4[ 0.007] | 1.5660 | 112 | -1.5 | 1 |
| 824.4 | 0.8 [ 0.02] | 824.57 | 79 | | 79 | 1.5 | 0.1[ 0.002] | 1.4450 | 110 | -1.5 | 1 |
| 824.5 | 0.8 [ 0.02] | 824.60 | 98 | | 98 | 3.6 | 0.4[ 0.004] | 3.5600 | 113 | 1.0 | 0 |
| 829 | 1 [ 0.003] | 829.40 | 160 | | 160 | 0.10 | 0.01[0.0001] | 0.0960 | 113 | -2.0 | 1 |
| 837 | 1 [ 0.002] | 836.55 | 165 | | 165 | 0.19 | 0.02[0.0001] | 0.1935 | 111 | -2.0 | 1 |
| 841.1 | 0.1 [ 0.08] | 841.70 | 100 | 10 [ 0.04] | 98 | 88 | 3 [ 2 ] | 76.6670 | 113 | 1.0 | 0 |
| 844 | 1 [ 0.02] | 843.53 | 165 | | 165 | 0.39 | 0.04[0.0002] | 0.3876 | 111 | -2.0 | 1 |
| 851.4 | 0.1 [ 0.04] | 851.70 | 120 | 10 [ 0.6] | 125 | 548 | 8 [ 7 ] | 534.6700 | 113 | 1.0 | 0 |
| 858 | 1 [ 0.03] | 858.51 | 165 | | 165 | 0.68 | 0.07[0.0004] | 0.6790 | 111 | -2.0 | 1 |
| 860.6 | 0.3 [ 0.2] | 861.31 | 160 | 20 [ 0.03] | 161 | 57 | 5 [ 0.6] | 59.8000 | 111 | 0.0 | 0 |

**Table 5 (cont.)– Resonance parameters for cadmium compared with ENDF/B-VII.1 parameters. Two uncertainties are given for each parameter, the Bayesian uncertainty from the SAMMY fit and an external error, in brackets, which conveys the agreement among individual sample fits.**

| E, eV | ΔE Bayesian [external] | Eendf | $\Gamma_\gamma$ | $\Delta \Gamma_\gamma$ Bayesian [external] | $\Gamma_\gamma$endf | $\Gamma_n$ | $\Delta \Gamma_n$ Bayesian [external] | $\Gamma_n$endf | isotope | $J^\pi$ | $\ell$ |
|---|---|---|---|---|---|---|---|---|---|---|---|
| 871 | 1 [ 0.03] | 870.60 | 160 | | 160 | 0.69 | 0.07[0.0006] | 0.6960 | 113 | -2.0 | 1 |
| 877.0 | 0.8 [ 0.2] | 876.48 | 165 | | 165 | 2.3 | 0.2[ 0.004] | 2.2530 | 111 | -2.0 | 1 |
| 878.1 | 0.2 [ 0.1] | 878.71 | 160 | 20 [ 0.02] | 161 | 43 | 3 [ 0.5] | 41.5000 | 111 | 1.0 | 0 |
| 881.8 | 0.6 [ 0.2] | 882.54 | 130 | | 130 | 8.6 | 0.8[ 0.02] | 8.5600 | 111 | 1.0 | 0 |
| 884.5 | 0.2 [ 0.05] | 884.84 | 90 | | 90 | 5.4 | 0.4[ 0.08] | 5.9010 | 112 | -1.5 | 1 |
| 888.7 | 0.2 [ 0.05] | 889.23 | 72 | 7 [ 0.04] | 73 | 53 | 3 [ 0.9] | 53.6800 | 116 | 0.5 | 0 |
| 895.0 | 0.4 [ 0.04] | 895.03 | 90 | | 90 | 2.8 | 0.3[ 0.02] | 2.8680 | 112 | -1.5 | 1 |
| 900 | 1 [ 0.03] | 900.44 | 165 | | 165 | 0.69 | 0.07[ 0.002] | 0.6790 | 111 | -2.0 | 1 |
| 901.1 | 0.3 [ 0.1] | 901.40 | 100 | 10 [ 0.05] | 98 | 67 | 5 [ 2 ] | 58.4000 | 113 | 0.0 | 0 |
| 903.9 | 0.4 [ 0.2] | 904.25 | 130 | | 130 | 12 | 1 [ 0.2] | 12.0300 | 111 | 1.0 | 0 |
| 908.7 | 0.1 [ 0.03] | 909.22 | 72 | 7 [ 0.4] | 69 | 274 | 6 [ 2 ] | 277.0000 | 112 | 0.5 | 0 |
| 911 | 1 [ 0.009] | 911.42 | 165 | | 165 | 0.68 | 0.07[0.0001] | 0.6790 | 111 | -2.0 | 1 |
| 912.6 | 0.3 [ 0.1] | 911.70 | 100 | 10 [ 0.02] | 98 | 75 | 7 [ 0.4] | 80.0000 | 113 | 0.0 | 0 |
| 917.4 | 0.3 [ 0.1] | 917.60 | 70 | | 70 | 8.8 | 0.7[ 0.1] | 8.5400 | 110 | -1.5 | 1 |
| 920 | 1 [ 0.02] | 919.70 | 160 | | 160 | 0.81 | 0.08[0.0003] | 0.8080 | 113 | -2.0 | 1 |
| 920.7 | 0.1 [ 0.04] | 921.18 | 83 | 8 [ 0.04] | 83 | 75 | 3 [ 1 ] | 66.8100 | 110 | 0.5 | 0 |
| 925.7 | 0.9 [ 0.1] | 925.31 | 165 | | 165 | 2.2 | 0.2[ 0.005] | 2.2390 | 111 | -2.0 | 1 |
| 929.3 | 0.9 [ 0.1] | 928.72 | 165 | | 165 | 1.9 | 0.2[ 0.002] | 1.9260 | 111 | -2.0 | 1 |
| 939.3 | 0.7 [ 0.1] | 939.70 | 100 | 10 [ 0.01] | 98 | 21 | 2 [ 0.09] | 20.4000 | 113 | 0.0 | 0 |
| 961 | 1 [ 0.08] | 961.33 | 165 | | 165 | 0.88 | 0.09[0.0003] | 0.8740 | 111 | -2.0 | 1 |
| 962.5 | 0.3 [ 0.08] | 962.39 | 72 | | 72 | 5.2 | 0.5[ 0.03] | 5.5150 | 114 | -1.5 | 1 |
| 966 | 1 [ 0.1] | 965.50 | 98 | | 98 | 18 | 2 [ 0.07] | 17.8670 | 113 | 1.0 | 0 |
| 964.8 | 0.4 [ 0.1] | 965.79 | 160 | 20 [ 0.06] | 158 | 60 | 4 [ 0.8] | 58.6800 | 111 | 1.0 | 0 |
| 970 | 1 [ 0.02] | 970.31 | 165 | | 165 | 0.58 | 0.06[0.0002] | 0.5820 | 111 | -2.0 | 1 |
| 974.9 | 0.5 [ 0.2] | 976.00 | 98 | | 98 | 12 | 1 [ 0.04] | 12.1330 | 113 | 1.0 | 0 |
| 993 | 1 [ 0.05] | 992.28 | 165 | | 165 | 1.4 | 0.1[ 0.001] | 1.3640 | 111 | -2.0 | 1 |
| 997 | 1 [ 0.07] | 998.00 | 160 | | 160 | 1.3 | 0.1[ 0.002] | 1.2720 | 113 | -2.0 | 1 |

Notes:

All negative energy resonances were taken from JENDL-4.0 and not varied.
All resonance parameters are reported to the precision of the more conservative of the two uncertainty estimates, Bayesian or external, with the exception of the neutron width for the 0.178 eV resonance.
Uncertainties for the 0.178 eV resonance were determined from thermal capture and transmission data.

Resonances at 4.5, 6.95, and 7.0 eV were fitted to a single thick capture sample, Cdtc20a. The results were inconclusive except for the resonance energies and the 4.5 eV resonance's neutron width.
Uncertainties for resonances from 18-600 eV were determined from epithermal capture and transmission data.
Uncertainties for resonances above 600 eV were determined from epithermal transmission data.

**Table 6- Average radiation widths in meV, derived from sensitive resonances according to the criterion $\Gamma_\gamma/\Gamma_n<5$ or $E < 1$ eV. Spin groups are separated where enough sensitive resonances are present to derive a meaningful average value.**

| Isotope/ spin All s-wave | Average $\Gamma_\gamma$, meV | s.d of distribution of $\Gamma_\gamma$, meV | Degrees of freedom (DF) of distribution of $\Gamma_\gamma$ from DF=2/var($\Gamma_\gamma$) | Number of resonances under 1 keV | Number of sensitive resonances under 1 keV |
|---|---|---|---|---|---|
| 106 | 120 | Too few resonances | Too few resonances | 2 | 2 |
| 108 | 140 | Too few resonances | Too few resonances | 2 | 2 |
| 110 | 90 | 30 | 20 | 5 | 4 |
| 111 J=0 | 160 | Too few resonances | Too few resonances | 2 | 1 |
| 111 J=1 | 170 | 40 | 35 | 33 | 11 |
| 112 | 62 | 8 | 110 | 5 | 4 |
| 113 J=0 | 103 | 6 | 512 | 11 | 5 |
| 113 J=1 | 130 | 30 | 36 | 28 | 13 |
| 114 | 59 | 7 | 135 | 4 | 4 |
| 116 | 72 | Too few resonances | Too few resonances | 2 | 1 |

**Table 7- Infinitely dilute capture resonance integral in barns for the measured Cd isotopes and from various evaluations.**

| Isotope | Mughabghab (1981) | Mughabghab Atlas (2006) | JENDL -4.0 | JEFF -3.2 | ENDF/B -VI.8 | ENDF/B -VII.0 | ENDF/B -VII.1 | NNL/RPI |
|---|---|---|---|---|---|---|---|---|
| 110 | 37 | 39 ± 4 | 39 | 33 | 41 | 39 | 33 | 34 ± 1 |
| 111 | 50 | 41.7 ± 3.0 | 45 | 51 | 50 | 47.4 | 51 | 50.6 ± 0.2 |
| 112 | 12 | 12.5 ± 1.0 | 13.1 | 13.3 | 13.5 | 13.3 | 13.2 | 13.0 ± 0.1 |
| 113 | 390 | 390 | 387 | 400 | 391 | 392 | 384 | 383 ± 1 |
| 114 | 23 ± 2 | 12.6 ± 1.0 | 16.8 | 12.5 | 13.1 | 13.1 | 12.5 | 12.6 ± 0.1 |

**Table 8- Total thermal cross sections in barns for the measured Cd isotopes and from various evaluations.**

| Isotope | Mughabghab (1981) | Mughabghab Atlas (2006) | JENDL -4.0 | JEFF -3.2 | ENDF/B -VI.8 | ENDF/B -VII.0 | ENDF/B -VII.1 | NNL/RPI |
|---|---|---|---|---|---|---|---|---|
| 110 | 11 ± 1 | 11 ± 1 | 11 | 11 | 11 | 11 | 11 | 10.9 ± 0.1 |
| 111 | 24 ± 3 | 6.9 ± 0.08 | 7.1 | 6.8 | 24 | 6.9 | 6.9 | 7.3 ± 0.1 |
| 112 | 2.2 ± 0.5 | 2.2 ± 0.5 | 2.1 | 2.2 | 2.2 | 2.2 | 2.2 | 2.2 ± 0.1 |
| 113 | 20600 ± 400 | 20615 ± 400 | 20062 | 20165 | 20615 | 20615 | 19862 | 20051 ± 14 |
| 114 | 0.30 ± 0.02 | 0.330 ± 0.018 | 0.34 | 0.34 | 0.34 | 0.34 | 0.31 | 0.2 ± 0.1 |

**Table 9- Nuclear radii for s-wave resonances from the current measurement and various nuclear libraries. The data were insensitive to the nuclear radii for p-wave resonances. Units are fermis, $10^{-13}$ cm.**

| Isotope | Natural abundance | Mughabghab (1981) | Mughabghab Atlas (2006) | JENDL-4.0 | JEFF-3.2 | ENDF/B-VI.8 | ENDF/B-VII.0 | ENDF/B-VII.1 | NNL/RPI |
|---|---|---|---|---|---|---|---|---|---|
| 106 | 0.0125 | NA | 6.33 | 6.2 | 6.00 | 5.7441 | 6.00 | 6.00 | 5.98 |
| 108 | 0.0089 | NA | 5.79 | 6.2 | 5.79 | 5.7441 | 5.79 | 5.79 | 5.78 |
| 110 | 0.1249 | NA | 5.5 | 6.2 | 5.80 | 6.418 | 5.80 | 5.80 | 5.59 |
| 111 | 0.1280 | NA | 5.9 | 6.2 | 6.15 | 5.7967 | 6.15 | 6.15 | 5.97 |
| 112 | 0.2413 | NA | 5.86 | 6.2 | 6.60 | 5.814 | 6.60 | 6.60 | 6.30 |
| 113 | 0.1222 | NA | NA | 6.2 | 6.50 | 6.7372 | 6.30 | 6.30 | 6.03 |
| 114 | 0.2873 | NA | NA | 6.2 | 6.90 | 6.31 | 6.90 | 6.90 | 6.58 |
| 116 | 0.0749 | NA | 6.19 | 6.2 | 6.24 | 6.50 | 6.24 | 6.24 | 6.16 |

**Table 10- Neutron s-wave strength functions in units of $10^{-4}$ for the most abundant Cd isotopes. The strength function is defined as the average reduced neutron width divided by the average level spacing. The results presented here were computed from the slope of the cumulative distribution of reduced neutron widths as described in Appendix A.**

| Cd isotope | Mughabghab (1981)[25] | Mughabghab Atlas (2006)[20] | JENDL-4.0 | JEFF-3.2 | ENDF/B-VI.8 | ENDF/B-VII.0 | ENDF/B-VII.1 | NNL/RPI |
|---|---|---|---|---|---|---|---|---|
| 110 | 0.44 ± 0.11 | 0.35 ± 0.07 | 0.43 ± 0.13 | 0.37 ± 0.13 | 0.42 ± 0.11 | 0.41 ± 0.12 | 0.37 ± 0.13 | 0.37 ± 0.13 |
| 111 | 0.8 ± 0.2 | 0.43 ± 0.07 | 0.55 ± 0.20 | 0.41 ± 0.11 | 0.39 ± 0.10 | 0.38 ± 0.09 | 0.41 ± 0.12 | 0.40 ± 0.12 |
| 112 | 0.5 ± 0.1 | 0.57 ± 0.11 | 0.51 ± 0.13 | 0.48 ± 0.15 | 0.52 ± 0.14 | 0.49 ± 0.15 | 0.48 ± 0.15 | 0.48 ± 0.15 |
| 113 | 0.31 ± 0.07 | 0.44 ± 0.06 | 0.35 ± 0.07 | 0.33 ± 0.07 | 0.33 ± 0.10 | 0.35 ± 0.06 | 0.35 ± 0.06 | 0.35 ± 0.06 |

**Table 11- The approximate maximum energy in keV of the region of linearity from the "staircase" plots (Figure 7) of number of resonances vs. resonance energy. These regions were used to calculate the s-wave strength functions given in Table 10.**

| Cd isotope | Mughabghab (1981)[25] | Mughabghab Atlas (2006)[20] | JENDL-4.0 | JEFF-3.2 | ENDF/B-VI.8 | ENDF/B-VII.0 | ENDF/B-VII.1 | NNL/RPI |
|---|---|---|---|---|---|---|---|---|
| 110 | Not specified in reference | Not specified in reference | 10 | 10 | 10 | 10 | 10 | 10 |
| 111 | Not specified in reference | Not specified in reference | 2.3 | 2.3 | 2.3 | 2.3 | 2.3 | 2.3 |
| 112 | Not specified in reference | Not specified in reference | 7 | 7 | 7 | 7 | 7 | 7 |
| 113 | Not specified in reference | Not specified in reference | 4 | 5 | 2.3 | 5 | 5 | 5 |

**Table 12-** The number of resonances in the region of linearity from the "staircase" plots (Figure 7) of number of resonances vs. resonance energy. These resonances were used to calculate the s-wave strength functions given in Table 10.

| Cd isotope | Mughabghab (1981)[25] | Mughabghab Atlas (2006)[20] | JENDL-4.0 | JEFF-3.2 | ENDF/B-VI.8 | ENDF/B-VII.0 | ENDF/B-VII.1 | NNL/RPI |
|---|---|---|---|---|---|---|---|---|
| 110 | Not specified in reference | Not specified in reference | 51 | 31 | 54 | 41 | 31 | 31 |
| 111 | Not specified in reference | Not specified in reference | 109 | 80 | 73 | 60 | 79 | 79 |
| 112 | Not specified in reference | Not specified in reference | 42 | 24 | 38 | 24 | 24 | 24 |
| 113 | Not specified in reference | Not specified in reference | 125 | 131 | 37 | 155 | 155 | 155 |

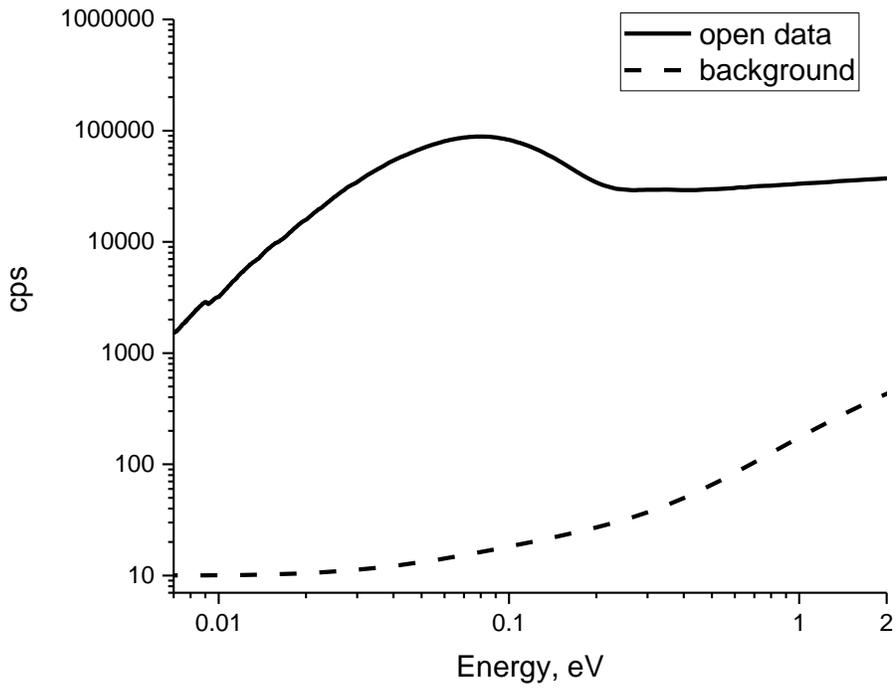

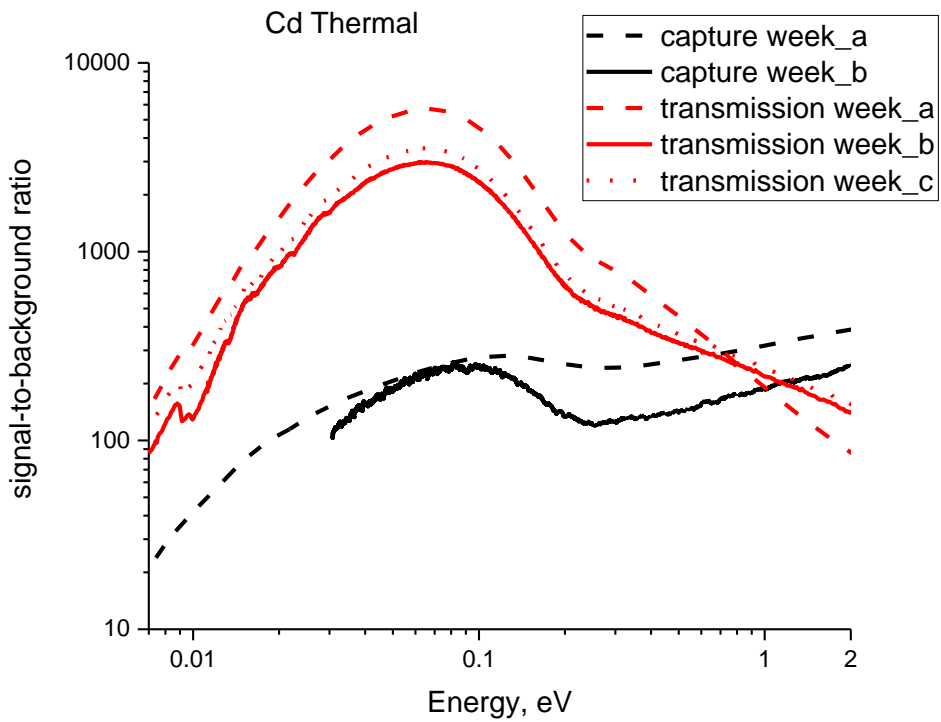

**Figure 1-** Cd thermal transmission signal and background counting rates (top) in counts per second (cps) from week_a, and signal-to-background ratios (bottom) for all thermal experiments.

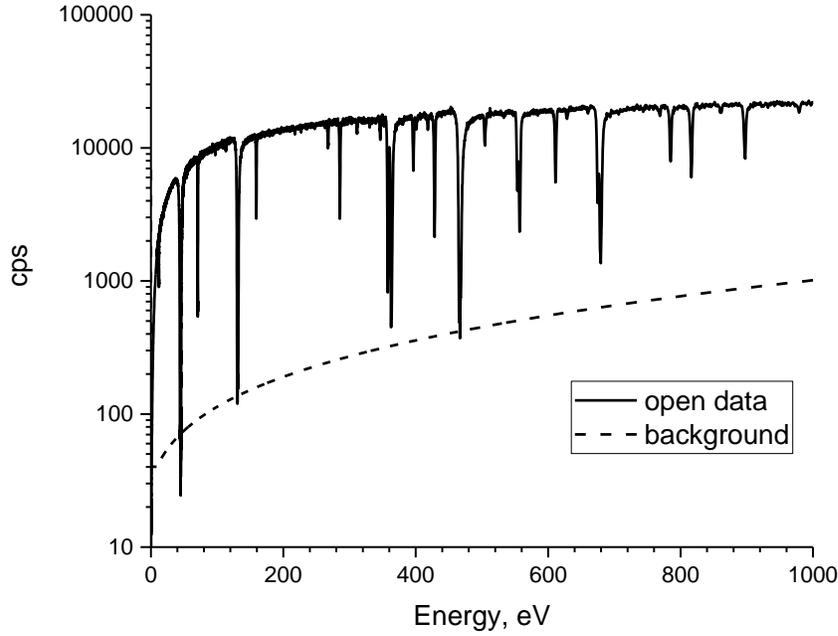

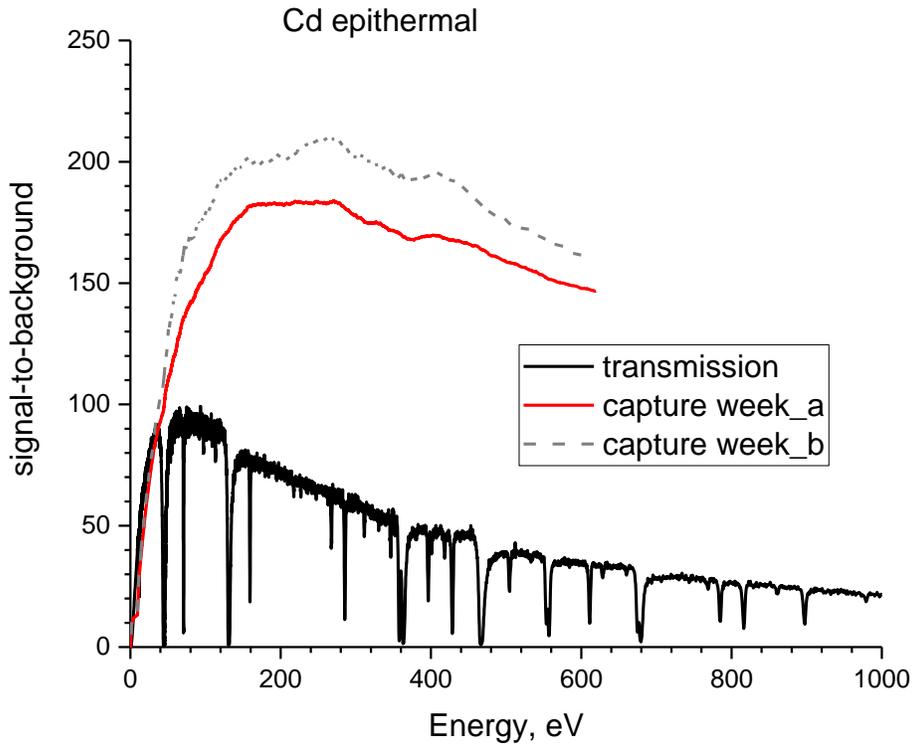

**Figure 2-** Cd epithermal transmission signal (open data, no sample) including the fixed Mo notch (accounts for resonance structure) and background (top), and signal-to-background ratios for all epithermal experiments (bottom).

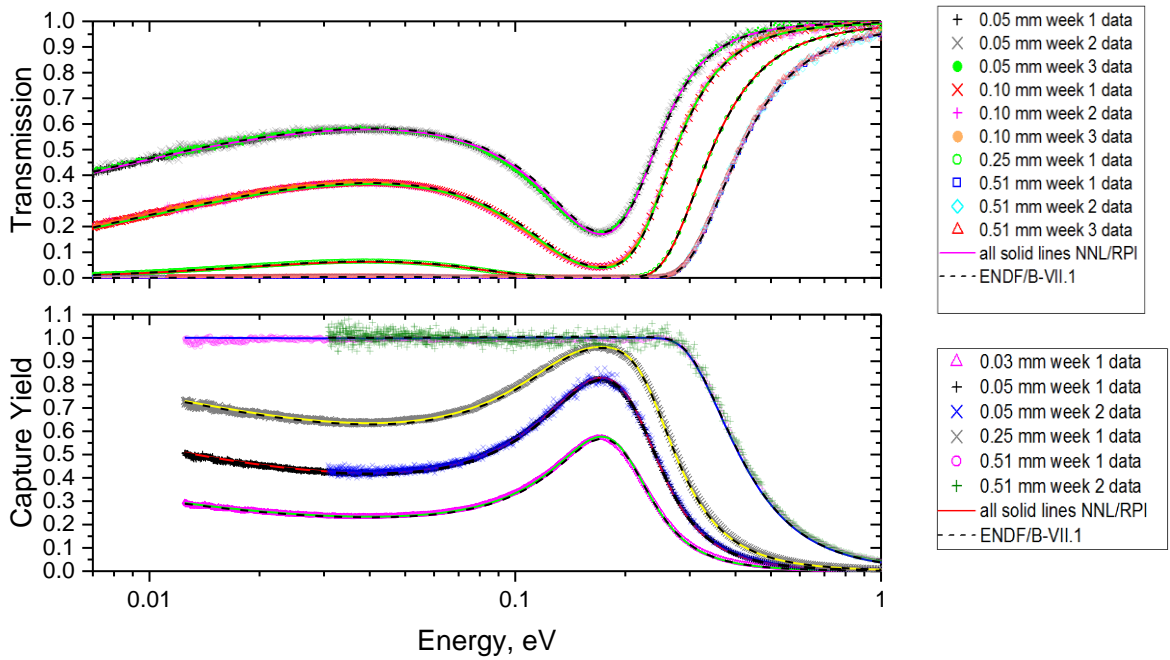

**Figure 3-** Overview of the thermal Cd transmission (top) and capture data (bottom) and curves calculated using NNL/RPI and ENDF/B-VII.1 resonance parameters. The samples are natural Cd. The thermal value is dictated by the 0.178 eV resonance in Cd113.

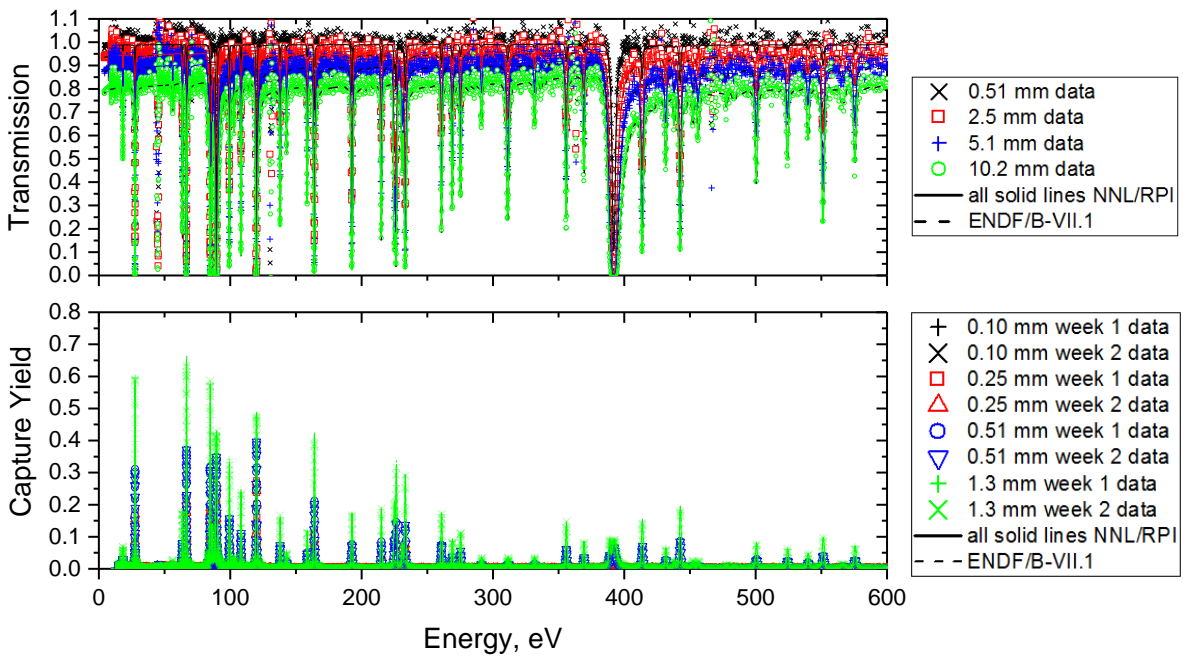

**Figure 4-** Overview of the epithermal Cd transmission (top) and capture data (bottom) in the 0-600 eV energy range and curves calculated using NNL/RPI and ENDF/B-VII.1 resonance parameters. The samples are natural Cd.

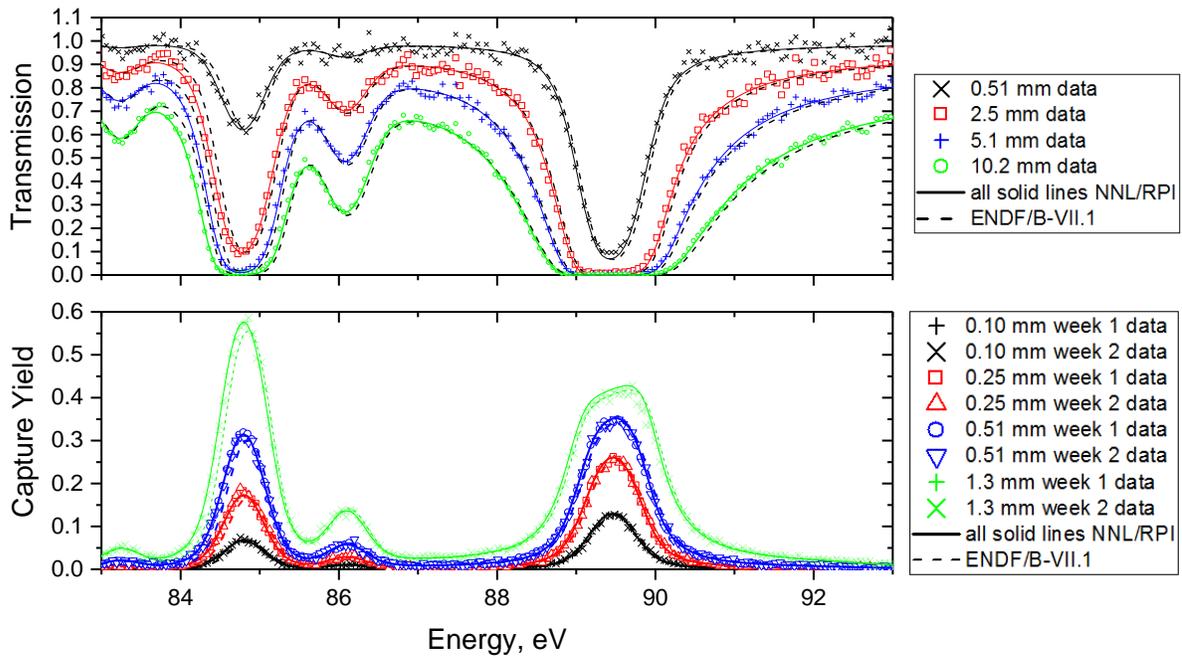

**Figure 5-** Epithermal Cd transmission (top) and capture data (bottom) in the 83-93 eV energy region and curves calculated using NNL/RPI and ENDF/B-VII.1 resonance parameters. The samples are natural Cd.

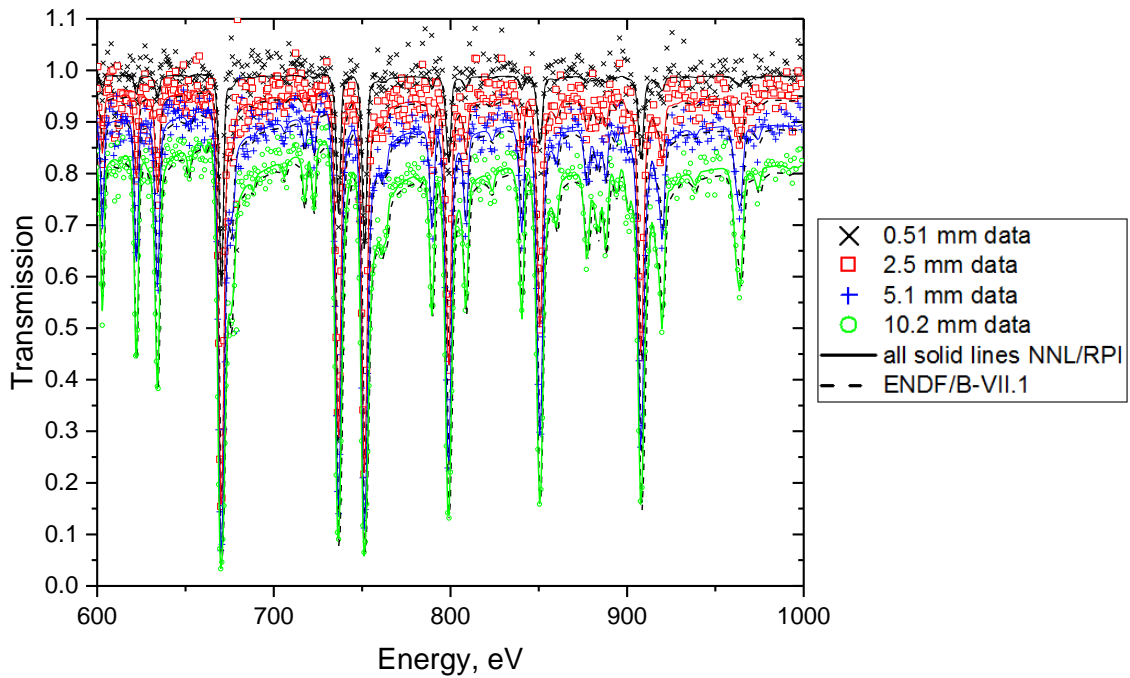

**Figure 6-** Epithermal Cd transmission data in the 600-1000 eV energy region and curves calculated using NNL/RPI and ENDF/B-VII.1 resonance parameters. The samples are natural Cd.

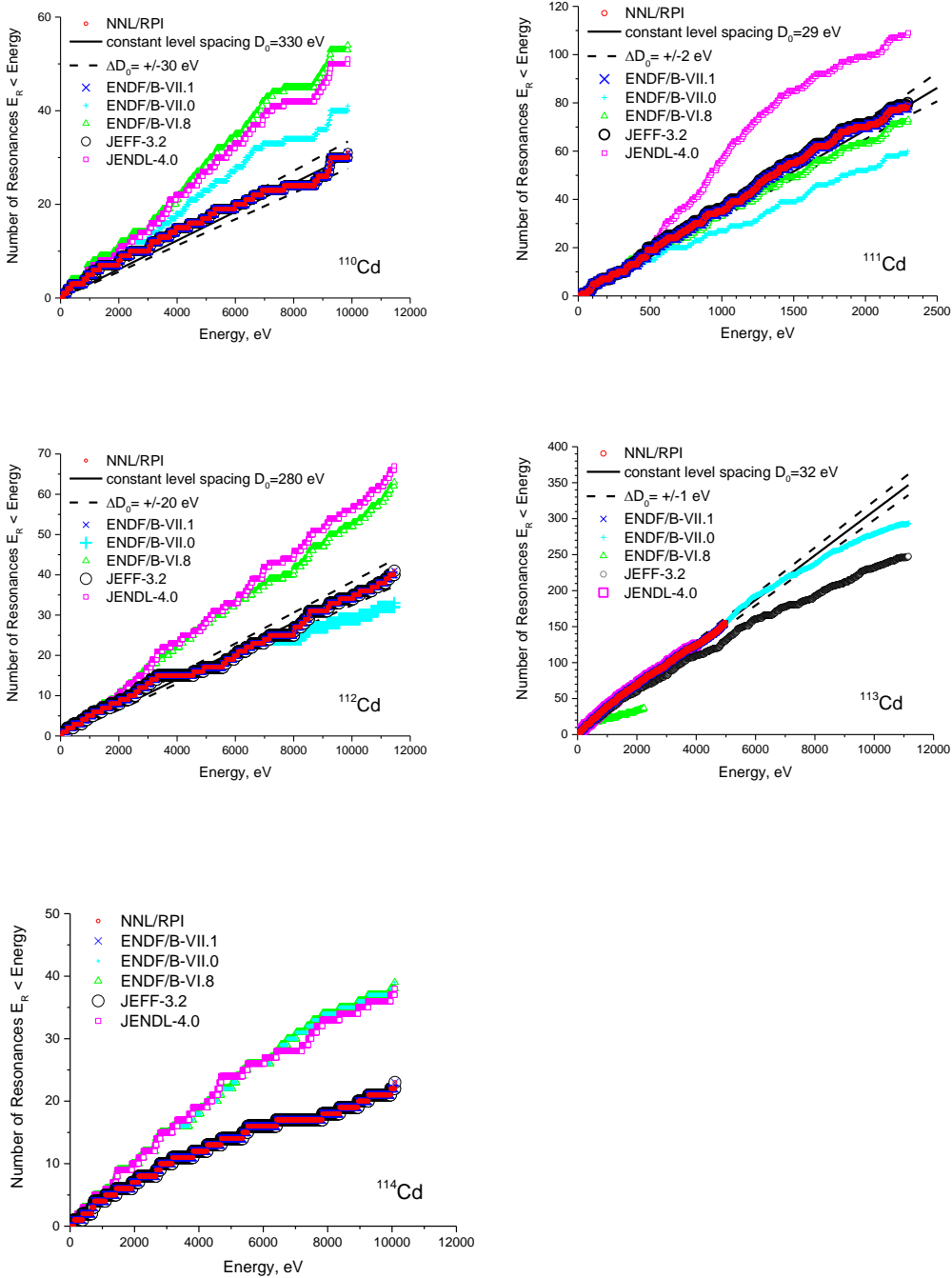

**Figure 7- Level density "staircase" plots for Cd isotopes of mass number 110-114. The level density given in each legend is the inverse slope of the best fit line up to the energy cutoff for linearity of level density for each isotope as given in Table 11. The number of resonances in the linear range for Cd110-113 is given in Table 12. No level density is estimated for Cd114 because there is no clear region of linearity.**

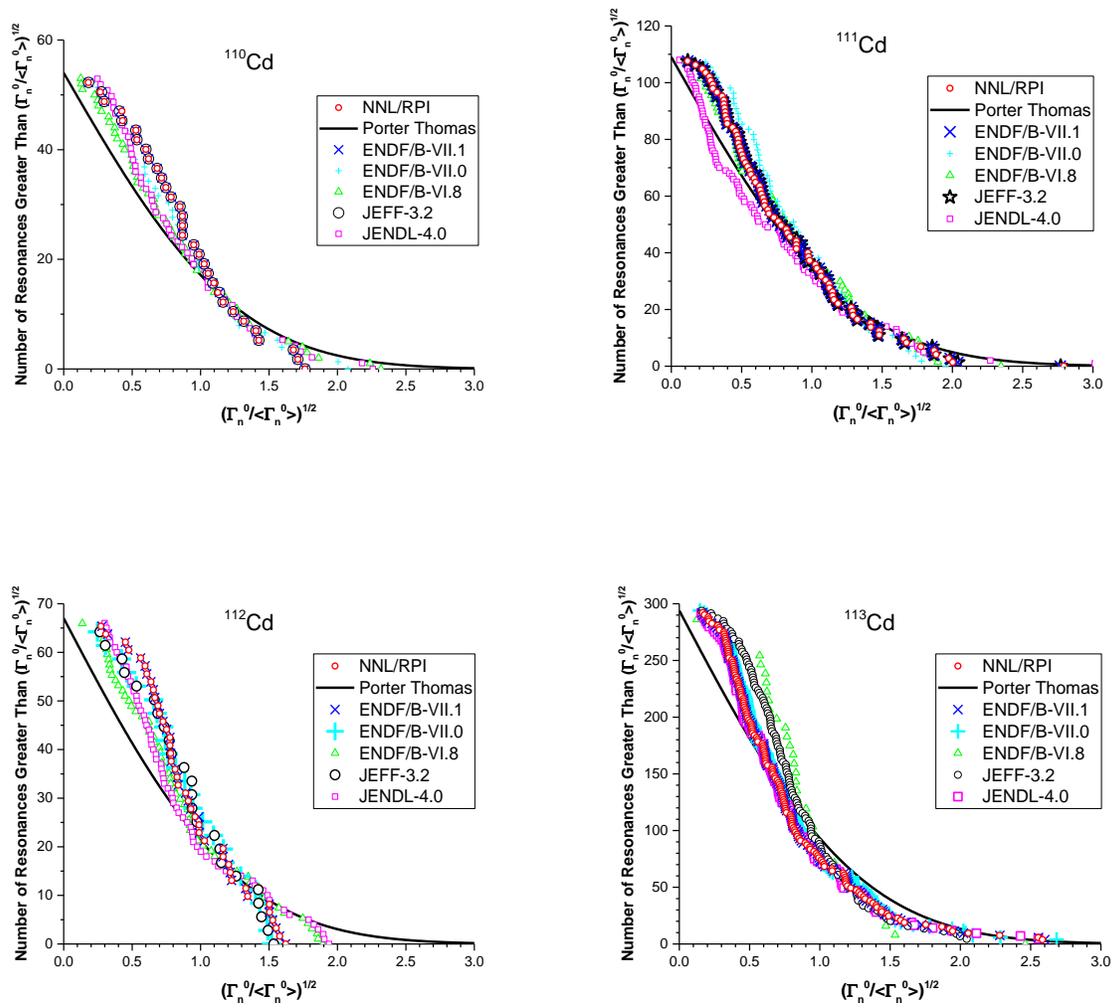

**Figure 8-** Reduced neutron width distributions compared with the theoretically-expected Porter Thomas distribution. Cd114 is excluded based on the uncertainty in the slope of the level density "staircase" plot, see Figure 7. The energy range for each plot is given in Table 11. The points in each plot in this figure were normalized such that every library for each isotope appears to have the same total number of resonances. The number of resonances used in each plot is given in Table 12.

**REFERENCES**

1. S. KOPECKY, I. IVANOV, M. MOXON, P. SCHILLEBEECKX, P. SIEGLER, I. SIRAKOV, "The total cross section and resonance parameters for the 0.178 eV resonance of $^{113}$Cd," *Nucl. Instrum. Meth. B*, **267**, pp. 2345–2350 (2009).
2. G. LEINWEBER, D. P. BARRY, J. A. BURKE, M. J. RAPP, R. C. BLOCK, Y. DANON, J. A. GEUTHER, and F.J. SAGLIME III, "Europium resonance parameters from neutron capture and transmission measurements in the energy range 0.01–20 MeV," *Ann. Nucl. Energy*, **69**, pp. 74 - 89, (2014).
3. R. C. BLOCK, M. C. BISHOP, D. P. BARRY, G. LEINWEBER, R. V. BALLAD, J. A. BURKE, M. J. RAPP, Y. DANON, A. YOUMANS, N. J. DRINDAK, G. N. KIM, Y.-R. KANG, M. W. LEE and S. LANDSBERGER, "Neutron transmission and capture measurements and analysis of Dy from 0.01 to 550 eV," *Prog. Nucl. Energ.*, **94**, pp. 126-132, (2017).
4. B. E. EPPING, G. LEINWEBER, D. P. BARRY, M. J. RAPP, R. C. BLOCK, T. J. DONOVAN, Y. DANON, and S. LANDSBERGER, "Rhenium Resonance Parameters from Neutron Capture and Transmission Measurements in the Energy Range 0.01 eV – 1 keV," *Prog. Nucl. Energ.*, **99**, pp. 59-72, (2017).
5. G. LEINWEBER, D. P. BARRY, R. C. BLOCK, J. A. BURKE, K. E. REMLEY, M. J. RAPP, and Y. DANON, "Neutron Capture and Total Cross Section Measurements of Cadmium at the RPI LINAC," *Proc. 13th International Topical Meeting on the Nuclear Applications of Accelerators*, July 31-August 4 2017, Quebec, Canada.
6. Y. DANON, R. E. SLOVACEK, and R. C. BLOCK, "The Enhanced Thermal Neutron Target at the RPI LINAC," *T. Am. Nucl. Soc.*, **68**, 473 (1993).
7. Y. DANON, R. E. SLOVACEK, and R. C. BLOCK, "Design and Construction of a Thermal Neutron Target for the RPI LINAC," *Nucl. Instrum. Meth. A*, **352**, 596 (1995).
8. M. E. OVERBERG, B. E. MORETTI, R. E. SLOVACEK, R. C. BLOCK, "Photoneutron Target Development for the RPI Linear Accelerator," *Nucl. Instrum. Meth. A*, **438**, 253 (1999).
9. Kamis Incorporated, Certification No. 005009-A, Mahopac Falls, New York.
10. R. C. BLOCK, P. J. MARANO, N. J. DRINDAK, F. FEINER, K. W. SEEMANN, and R. E. SLOVACEK, "A Multiplicity Detector for Accurate Low-Energy Neutron Capture Measurements," *Proc. Int. Conf. Nuclear Data for Science and Technology*, May 30-June 3, 1988, Mito, Japan, p. 383.
11. D. P. BARRY, "Neodymium Neutron Transmission and Capture Measurements and Development of a New Transmission Detector," PhD Thesis, Rensselaer Polytechnic Institute (2003).
12. Y. DANON, "Design and Construction of the RPI Enhanced Thermal Neutron Target and Thermal Cross Section Measurements of Rare Earth Isotopes," PhD Thesis, Rensselaer Polytechnic Institute (1993).
13. N. M. LARSON, "Updated Users' Guide for SAMMY: Multilevel R-Matrix Fits to Neutron Data Using Bayes' Equations," ORNL/TM-9179/R8 ENDF-364/R2, Oak Ridge National Laboratory (October 2008).
14. M. B. CHADWICK et al., "ENDF/B-VII.1: Nuclear Data for Science and Technology: Cross Sections, Covariances, Fission Product Yields and Decay Data," *Nucl. Data Sheets*, **112**, 2887 (2011).
15. Y. DANON and R. C. BLOCK, "Minimizing the Statistical Error of Resonance Parameters and Cross Sections Derived from Transmission Measurements," *Nucl. Instrum. Meth. A*, **485**, 585 (2002).
16. K. SHIBATA, O. IWAMOTO, T. NAKAGAW, N. IWAMOTO, A. ICHIHARA, S. KUNIEDA, S. CHIBA, K. FURUTAKA, N. OTUKA, T. OHASAWA, T. MURATA, H. MATSUNOBU, A. ZUKERAN, S. KAMADA, and J.-I. KATAKURA, *J. Nucl. Sci. Technol.*, **48**, 1 (2011).
17. Y. DANON, D. WILLIAMS, R. BAHRAN, E. BLAIN, B. McDERMOTT, D. BARRY, G. LEINWEBER, R. BLOCK, and M. RAPP, "Simultaneous Measurements of $^{235}$U Fission and Capture Cross Sections From 0.01 eV to 3 keV Using a Gamma Multiplicity Detector," *Nucl. Sci. Eng.*, **187**, Iss. 3, (2017).
18. G. LEINWEBER, D. P. BARRY, J. A. BURKE, N. J. DRINDAK, Y. DANON, R. C. BLOCK, N. C. FRANCIS, and B. E. MORETTI, "Resonance Parameters and Uncertainties Derived from Epithermal Neutron Capture and Transmission Measurements of Natural Molybdenum," *Nucl. Sci. Eng.*, **164**, 287 (2010).
19. M. J. TRBOVICH, "Hafnium Neutron Cross Sections and Resonance Analysis," PhD Thesis, Rensselaer Polytechnic Institute (2003).
20. S. F. MUGHABGHAB, *Atlas of Neutron Resonances*, 5th ed., Elsevier, New York (2006).
21. A. SANTAMARINA, D. BERNARD, P. BLAISE, M. COSTE, A. COURCELLE, T. D. HUYNH, C. JOUANNE, P. LECONTE, O. LITAIZE, S. MENGELLE, G. NOGUERE, J.-M. RUGGIERI, O. SEROT, J. TOMMASI, C. VAGLIO, and J.-F. VIDAL, "The JEFF-3.1.1 Nuclear Data Library," JEFF Report 22 (Nuclear Energy Agency Organisation for Economic Co-operation and Development, 2009).

22. R. E. MacFARLANE and D. W. MUIR, "The NJOY Nuclear Data Processing System Version 91," LA-12740-M, Los Alamos National Laboratory (1994).
23. C. L. DUNFORD, "ENDF Utility Codes Release 6.12," Informal Report (2001).
24. P. F. ROSE and C. L. DUNFORD, "ENDF-102 Data Formats and Procedures for the Evaluated Nuclear Data File ENDF-6," BNL-NCS-44945, Rev. 2, Brookhaven National Laboratory (1997).
25. S. F. MUGHABGHAB, R. R. KINSEY, C. L. DUNFORD, *Neutron Cross Sections* Vol. 1 Part A, Academic Press, Orlando, Florida (1981).
26. M.B. CHADWICK et al., "ENDF/B-VII.0: Next Generation Evaluated Nuclear Data Library for Science and Technology," *Nucl. Data Sheets*, **107**, 12, 2931 (2006).
27. C. E. PORTER and R. G. THOMAS, "Fluctuations of Nuclear Reaction Widths," *Phys. Rev.*, **104**, 483 (1956).
28. S. KOPECKY, I. IVANOV, M. MOXON, P. SCHILLEBEECKX, P. SIEGLER, I. SIRAKOV, "The total cross section and resonance parameters for the 0.178 eV resonance of $^{113}$Cd," *Nucl. Instrum. Meth. B*, **267**, pp. 2345–2350 (2009).
29. U. HARZ and H. G. PRIESMEYER, "The total neutron cross-section of Cd-113 at 0.178 eV," *Atomkernenergie*, **24,** 142 (1974).
30. F. WIDDER and J. BRUNNER, "Total cross section of cadmium for neutrons of energy between 0.01 and 10 eV," *Nukleonik*, **11**, 297 (1968).
31. R. O. AKYUZ, C. CANSOY, and F. DOMANIC, "Parameters for the first neutron resonance in $^{113}$Cd," *Nucl. Sci. Eng.*, **28**, 359 (1967).
32. Ju. SHCHEPKIN, Ju. ADAMCHUK, L. DANELJAN, G. MURADJAN, ''Neutron spectroscopic study of separated cadmium isotopes,'' *Nuclear data for reactors, Vol. I, Conference Proceedings*, Paris, 1966, IAEA, pp. 93–99.
33. E. MESERVEY, "Neutron-capture cross sections by capture-gamma counting," *Phys. Rev.*, **96**, pp. 1006–1013, (1954).
34. B. N. BROCKHOUSE, "Resonant scattering of slow neutrons," *Can. J. Phys.*, **31**, 432 (1953).
35. L. J. RAINWATER, W. W. HAVENS JR., C. S. WU, J. R. DUNNING, "Slow neutron velocity spectrometer studies I. Cd, Ag, Sb, Ir, Mn," *Phys. Rev.*, **71**, pp. 65–79, (1947).
36. L. J. RAINWATER, W. W. HAVENS JR., "Neutron beam spectrometer studies of boron, cadmium and the energy distribution from paraffin," *Phys. Rev.*, **70**, pp. 136–153, (1946).

**Appendix A    Estimation of the Strength Function**

Two methods for computing and estimating the s-wave strength function over an energy interval of interest are presented here. While both methods will lead to reasonably accurate estimates of the strength function, the second method discussed gives an analyst a clearer picture of the behavior of resonances as energy increases, with uncertainties that follow departures from linear behavior.

The reduced neutron widths for s-wave resonances, $\Gamma_n^0$, are defined in Eq.(A-1).

$$\Gamma_n^0 = \sqrt{\frac{1eV}{E}} \Gamma_n, \qquad (A-1)$$

where $\Gamma_n$ is the neutron width of the resonance and $E$ is the resonance energy in eV. In addition to reduced neutron width, statistical factor, $g$, is needed to compute strength functions. This statistical factor is defined by Eq.(A-2).

$$g = \frac{2J+1}{2(2I+1)}, \qquad (A-2)$$

where $J$ is the spin of the resonance and $I$ is the spin of the nucleus.

A.1 Estimation through Definition of Strength Function

Reference A.1 gives the definition of the s-wave strength function, $S_0$, as shown in Eq.(A-3),

$$S_0 = \frac{<g\Gamma_n^0>}{D_0}, \qquad (A-3)$$

where $<g\Gamma_n^0>$ is the average reduced neutron width multiplied by statistical factor, $g$. $D_0$ is the average s-wave level spacing, or equivalently, in Eq.(A-4),

$$S_0 = \frac{\sum g\Gamma_n^0}{\Delta E}, \qquad (A-4)$$

Where $\Sigma g\Gamma_n^0$ is the cumulative $g\Gamma_n^0$ over energy range of interest $\Delta E$.

Given the definition (A-3), the uncertainty in the strength function, $\Delta S_0$, is estimated through the error propagation formula as shown in Eq.(A-5), assuming that uncertainties in $<g\Gamma_n^0>$ and $D_0$ are independent,

$$\left(\frac{\Delta S_0}{S_0}\right)^2 = \left(\frac{\Delta <g\Gamma_n^0>}{<g\Gamma_n^0>}\right)^2 + \left(\frac{\Delta D_0}{D_0}\right)^2 , \quad (A-5)$$

where $\Delta<g\Gamma_n^0>$ is the standard error on $<g\Gamma_n^0>$, and $\Delta D_0$ is the uncertainty on the average level spacing determined using Eq.(A-6) by assuming, from Reference A.1, a Wigner distribution,

$$\Delta D_0 = \frac{0.52 D_0}{\sqrt{N}} , \quad (A-6)$$

where $N$ is the number of levels. Experience has shown that the uncertainty in strength function estimated with this method is dominated by the uncertainty in the level spacing.

A.2 Estimation through Least Squares Fitting

An alternative technique for computing strength function is employed as well. Plots of cumulative $g\Gamma_n^0$ against energy are made, called staircase plots in Reference A.1, and best-fit lines are determined based upon the accumulation $\Sigma g\Gamma_n^0$ via least-squares fitting. When plotted against energy, the accumulation is generally expected to be linear. An example of such linear behavior as well as a best-fit line of the accumulation is shown in Figure A-1.

Referring to Figure A-1 the coefficients $A$ and $B$ are obtained for the best fit line $y = AE + B$ for the accumulation versus energy, then the strength function can be computed from Eq.(A-7) as

$$S_0 = A = \frac{y(E_2) - y(E_1)}{E_2 - E_1} , \quad (A-7)$$

Where $E_1$ and $E_2$ are the lower and upper bounds of the energy region being investigated, respectively.

The reader will observe that this equation is simply the slope of the best-fit line. For the case of s-wave strength functions with one best-fit line over the entire energy range, this is strictly true. However, in future applications, more sophisticated computations will be performed. For instance, higher level (e.g., p-wave) strength functions can be computed by including $2l+1$ in the denominator of Eq. (A-7), where $l$ is the angular momentum quantum number of an incoming neutron. The above equation also holds in cases where different best fit lines are used in different energy regions to treat radical departures from linear behavior in staircase plots, as in Reference A.2. For these reasons, strength function is computed with Eq. (A-7).

The uncertainty in this strength function, $\Delta S_0$, is estimated as the quadrature sum of the least-squares fitting error, $\Delta S_0^{fit}$, and the standard deviation of strength function values about the central value computed with Eq. (A-7), $\Delta S_0^{std}$ in Eq.(A-8),

$$\Delta S_0 = \sqrt{\left(\Delta S_0^{fit}\right)^2 + \left(\Delta S_0^{std}\right)^2} \quad , \tag{A-8}$$

The least-squares fitting error is given in Eq.(A-9) as

$$\Delta S_0^{fit} = \frac{\sqrt{\Delta y(E_2)^2 + \Delta y(E_1)^2}}{E_2 - E_1} \quad , \tag{A-9}$$

where the uncertainty in the value at a point given by the best-fit line, $\Delta y(E)$, is estimated by Eq.(A-10),

$$\Delta y(E) = \sqrt{\left(E\Delta A\right)^2 + \left(\Delta B\right)^2 + 2E\Delta AB} \quad . \tag{A-10}$$

In Eq. (A-10) $\Delta A$ is the uncertainty in the slope of the best-fit line $y = AE + B$, $\Delta B$ is the uncertainty in the y-intercept, and $\Delta AB$ is the covariance between the two fitted parameters.

The standard deviation of strength function values is calculated from Eq.(A-11),

$$\Delta S_0^{std} = \sqrt{\frac{\sum_i (S_{0,i} - S_0)^2}{N - 1}} \quad , \tag{A-11}$$

where $S_0$ is the strength function computed with Eq.(A-7), $N$ is the total number of resonances in the energy range, and $S_{0,i}$ is the strength function computed using Eq.(A-5) for the $i$-th resonance in the interval, as given in Eq.(A-12),

$$S_{0,i} = \frac{\sum_{k=1}^{i} g\Gamma_{n,k}^0}{E_i - E_1} \quad . \tag{A-12}$$

In Eq.(A-12) $E_i$ is the energy of the *i*-th resonance in the interval, $E_1$ is the energy of the first resonance in the interval, and $\Gamma^0_{n,k}$ is the reduced neutron width for the *k*-th resonance in the interval $[E_i, E_1]$.

### A.3 Conclusions

Clearly, the method of strength function estimation given in Eq.s(A-7) through (A-12) gives an analyst much more information about the accumulation $\Sigma g\Gamma^0_n$ over energy, particularly departures from linear increases in $\Sigma g\Gamma^0_n$, than the first method presented, given in Eq.s (A-3) through (A-5). Because of this increased detail, it is recommended that an analyst use the resultant strength function and the uncertainty from the least-squares fit method over an estimation of strength function through its definition.

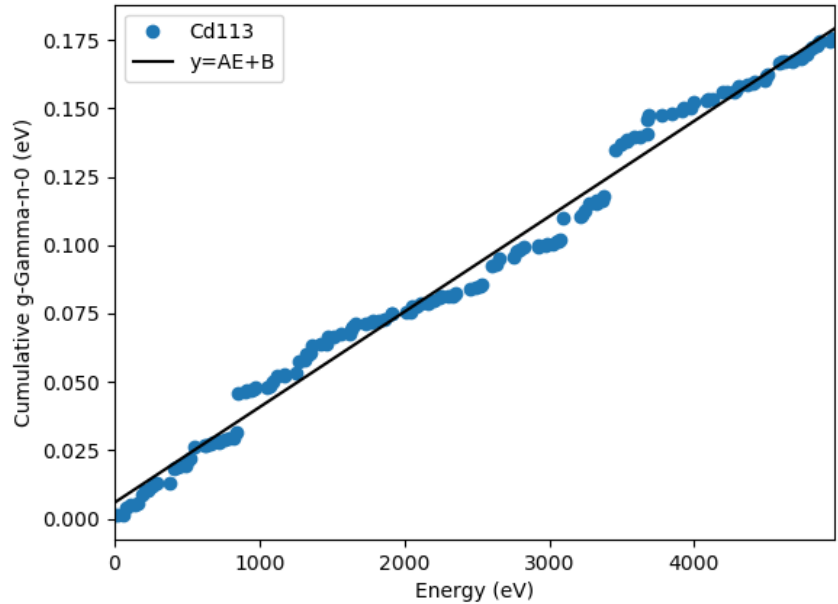

**Figure A-1. Sample plot of $\Sigma g\Gamma^0_n$ vs. *E* for the s-wave spin groups of Cd113. The black line represents the best fit ($A = S_0 = 0.35 \times 10^{-4}$, $B = 0.58 \times 10^{-2}$) of accumulation given the data.**

**References**

A.1 S. F. MUGHABGHAB, *Atlas of Neutron Resonances*, 5th ed., Elsevier, New York (2006).
A.2 R. E. CHRIEN et al., "Failure of Bohr's compound nucleus hypothesis for the $^{98}$Mo(n,γ)$^{99}$Mo reaction," *Phys. Rev. C*, **13** No. 2, pp. 578-594 (1976).